\documentclass[a4paper,11pt]{article}
\pdfoutput=1
\usepackage{jheppub} \usepackage[T1]{fontenc} 

\newcommand{\be}{\begin{equation}}
\newcommand{\ee}{\end{equation}}
\newcommand{\ba}{\begin{eqnarray}}
\newcommand{\ea}{\end{eqnarray}}

\title{Production of black holes and string balls in a two-dimensional split-fermion model}
\author{Shohreh Abdolrahimi ${}^a$}
\author{and Christos Tzounis ${}^{b,} \, {}^a$}
\affiliation{${}^a$ Institut f{\"u}r Physik, Universit{\"a}t Oldenburg, \\
Postfach 2503 D-26111 Oldenburg, Germany}
\affiliation{${}^b$ Theoretical Physics Institute, University of Alberta,\\
Edmonton, AB, Canada,  T6G 2G7}
\emailAdd{abdolrah@ualberta.ca}
\emailAdd{tzounis@ualberta.ca}

\abstract{We present cross-sections for the black hole and string ball production in
proton-proton collisions in a TeV-scale gravity model with split
fermions in two dimensions.      
The cross-section for black hole and string ball production in the
split-fermion model is smaller than in the non-split-fermion model. The drop of the cross-section for the string ball production can be one to two orders of magnitude with the increase of the width of the brane from $L=0$ to 15 $\text{TeV}^{-1}$. The cross-section for string ball production in two-dimensional split fermion model reduces more in comparison to black holes. Black holes are quite hard to be observed at the LHC. In fact, taking into account the current experimental limits on the fundamental Planck scale, black holes cannot be produced at the LHC. Cross-section for string ball production depends significantly on string coupling constant, making it very model dependent.
We investigate the range of values of string coupling constant from 0.02 to 0.4. There has been no evidence for production of string balls at $\sqrt{s}=8$ TeV. A two-dimensional split fermion model with a extremely thick brane can account for the absence of signature of string balls for a world with the value of fundamental Planck scale even as low as 1 TeV.  
}


\begin{document}
\maketitle

\flushbottom

\section{Introduction} \label{sec1}

Models of
large or
warped~\cite{1,2,3,4,5} extra dimensions allow the
fundamental scale of gravity to be as low as the electroweak scale, providing us with a low-scale gravity model. With a black hole solution finally discovered in Randall-Sundrum model \cite{6,7,8}, warped extra dimensions can have validation as well. 
One of the predictions of low-scale gravity models is the possibility of
black hole production in particle collisions with energies above the
gravity scale, such as in proton-proton collisions at the Large Hadron 
Collider (LHC) for TeV-scale gravity.  

Low-scale quantum gravity models have some unacceptable predictions,
such as fast proton decay, large $n\bar{n}$ oscillations, large mixing
between leptons, etc.  
A solution to some of the unacceptable predictions of low-scale quantum
gravity models is the split-fermion
model~\cite{9,10}. 
In this model, the Standard Model particles are confined to a ``thick''
brane extending in the higher dimensional bulk space.
In the simplest case, this thickness is one-dimensional, and quarks and
leptons live on different slices of this thick brane.
The split-fermion model also has some other advantages.
The model allows a geometric interpretation of the observed fermion
masses. 
Namely, the effective four-dimensional Yukawa couplings can be viewed as
overlaps of the wave functions between two different chiral fermions and
is lower as their relative distance in the split-fermion dimension
increases. In a one split-fermion dimension model, it is
not possible to obtain sufficient CP violation in accordance with the
experimental data. Considering two split-fermion dimensions, i.e., a six-dimensional
Standard Model~\cite{11}, one can find an example where all the quark masses and
the mixing angles, as well as the required strength of CP violation
in quark sector are reproduced. In order to make serious predictions, one needs to work within the context of realistic phenomenologically valid models. We consider that, a two-dimensional split fermion model is such a model for a TeV-scale quantum gravity.

The analysis of black hole production in colliders range from using a
simple $\pi r_g^2$ form for the cross-section, where $r_g$ is the
gravitational radius of the black hole formed in the parton scattering
process~\cite{12,13,14}, to
calculations based on classical general relativity using the
trapped-surface
approach~\cite{15,16,17}. 
The effects of finite particle
size~\cite{18,19}, angular
momentum~\cite{20,21,22,23,24}, 
and electric charge~\cite{25,26}
on the cross-section are sometimes included in limited form.    

If the black hole is formed with mass not too close to the Planck scale,
it is anticipated to decay by emitting Hawking
radiation~\cite{27}. If the black hole is produced at the LHC, it would decay to final states with a relatively high multiplicity. The theory of black hole evaporation is based on a semiclassical description involving quantum fields on classical space-time backgrounds. Although in the classical general relativity theory black holes can only absorb and not emit particles, in the semiclassical description black holes  create and emit particles as if they were hot bodies. 
The radiation from the black hole is thermal, meaning information would
be permanently lost inside the event horizon, and there would be no
$S$-matrix to take an initial pure state to a final pure state,
violating unitarity in quantum mechanics~\cite{28}. 
Although it has been suggested that some new dynamics provided by
quantum gravity at short distances would resolve this issue at the final
stage of the black hole evaporation, it is widely
accepted~\cite{29,30} now that most of
the information comes out with the bulk of the radiation to give an
$S$-matrix, allowing the number of possible microstates of the black
hole to be proportional to the exponential of its entropy.  
If so, the black hole could make a transition to a highly-excited string
state as it evaporates. 

This string theory motivated scenario provides a convincing picture of
the evolution of a black hole at the last stages of its evaporation.
As the black hole shrinks, it eventually reaches the ``correspondence
point'' $M_{min}\sim M_s/g_s^2$, (where $M_s$ is the string scale related to the fundamental Planck mass and $g_s$ is the dimensionless string coupling constant), and makes a transition to a configuration dominated by a
highly-excited long string called a string
ball \cite{31,32,33}.  
String balls evaporate thermally at the Hagedorn temperature and give rise to high-multiplicity events containing hard primary photons and charged leptons, which have negligible standard-model background \cite{34}.
The result is that, the evaporating black hole leaves no stable remnant. 

If the Planck scale is a few TeV, the mass of the black hole has to
be close to the maximum LHC energy for it to be described by general
relativity.
Because of the steeply falling parton distribution functions with
increasing mass, this leads to a black hole production cross-section
that is significantly reduced at the LHC. 
However, because of the correspondence between black holes and
string balls, black hole production is dual to the production of a
highly-excited long string state. 
Thus, if the string scale and the Planck scale are a few TeV in large
extra dimensions, we might expect TeV-scale string physics at the LHC.
String balls would be expected to be produced at the LHC much more than black 
holes~\cite{35,36,37,38}. 

Black holes and string balls have been searched for in proton-proton
collisions at the LHC. ATLAS collaboration has performed a search for an excess of events with multiple high transverse momentum
objects including charged leptons and jets, using 20.3 fb$^{-1}$ of proton-proton
collision data recorded by the ATLAS detector at the Large Hadron Collider in 2012 at
a centre-of-mass energy of $\sqrt{s}=8$~TeV \cite{39}. No excess of events beyond Standard Model expectations have been observed (also see \cite{40,41}). The CMS collaboration~\cite{42,43}
has published using up to 4.7~fb$^{-1}$ of 7~TeV data.
In addition, CMS has published a result using 3.7~fb$^{-1}$
of 8~TeV collision data \cite{44}.
In all cases, agreement with the expected Standard Model backgrounds
were obtained. The CMS experiment has excluded up to a black hole mass of 3.5-4.5 TeV in a variety of theoretical models. 

In this paper, we present cross-sections for the black hole and string ball production for
proton-proton collisions in a TeV-scale gravity model with split
fermions in two dimensions.
We compare the cross-section for black hole and string ball production
in particle collisions in the non-split-fermion model to a
two-dimensional split-fermion model.  
We illustrate that the split-fermion model affects the string ball cross
section more than the black hole cross-section. Given the lack of signature of black holes or string balls for the centre-of-mass energy $\sqrt{s}=8$ TeV at the LHC, limits on the value of the fundamental Planck scale can be imposed. These limits depend on the width of split fermion dimension and the value of the string coupling constant. We investigate the range of values of string coupling constant from 0.02 to 0.4. Furthermore, no evidence for production of string balls at $\sqrt{s}=8$ TeV suggests that if our world has extra dimensions with the fundamental scale of multidimensional gravity being between one to two TeV units then, it should be something like a two-dimensional split fermion model with a ``thick'' brane.

This paper is organized as follows:
In Sec. 2 we define the model we are considering, and in
Sec. 3 we present our numerical results, and their discussion. 
We have largely used the BlackMax Monte Carlo event
generator~\cite{45,46} for our computations. We present some of the details for running of the BlackMax code in Appendix. 

\section{Black holes and string balls} 

\subsection{Two-dimensional split-fermion model} 

In this paper, we consider both black hole and string ball production
cross-sections in a split-fermion model with two split-fermion
dimensions.
Two split-fermion dimensions helps solve some of the unacceptable
predictions of the low-scale quantum gravity models, as well as
providing us with an example where all the quark masses, the mixing
angles, and the required strength of CP violation in quark sector can be
reproduced \cite{11}.   

In the simplest split-fermion model \cite{47}, only one split-fermion dimension is considered,
which gives the five-dimensional Standard Model with the fermions being 
localized in the extra dimensional brane, while gauge bosons and Higgs
fields can propagate in the extra dimension.
If one assumes that fermions have a Gaussian distribution in the
fifth dimension, and one universal Yukawa coupling to the bulk 
Higgs field, one can find a unique configuration of the Standard Model
fermions positions which can explain the Yukawa mass hierarchies by
displacing the left-handed and right-handed components of the Standard
Model fermions without imposing new symmetries~\cite{47}.
This configuration fits all quark and lepton masses, and mixing angles
of the four-dimensional Standard Model. 
However, if one considers only one split-fermion dimension then, it is
not possible to obtain sufficient CP violation in accordance with the
experimental data while having one universal coupling to the bulk Higgs
field. 
Considering two split-fermion dimensions, i.e., a six-dimensional
Standard Model~\cite{11}, one can find an example where all the quark masses and
the mixing angles, as well as the required strength of CP violation
in quark sector are reproduced, even though this may not be a unique
configuration.
Note that, it is also possible to produce the required CP violation in
the five-dimensional Standard Model (one-dimensional split-fermion
model) by having different Yukawa couplings for up-type and down-type
quarks~\cite{48}.

We assume that Standard Model fermions have different positions in the
extra two split-fermion dimensions, and common Gaussian profiles about
these positions.
Thus, in the two-dimensional split-fermion model 

\begin{equation}\label{eq1}
\Psi(\vec{x}) = \sqrt{\frac{2}{\pi}} \mu \exp\left[ -\mu^2(x_5^2 +
x_6^2)\right] \psi(x_0,x_1,x_2,x_3)\, , 
\end{equation} where, $x_5$ and $x_6$ are the split-fermion dimensions, and $\psi$ is a 
canonically normalized massless left-handed four-dimensional fermion
wave function.   
For simplicity, we consider the same Gaussian width $\sqrt{2}/\mu$ for
both split-fermion dimensions $x_5$ and $x_6$.  
The location of the fermion wave functions are taken from
Ref.~\cite{11} and shown in Table~\ref{tab1}.

\begin{table}[htdp]
\centering
\begin{tabular}{|l|ccc|}\hline
Quark type           & $(u,d)_L$ & $(c,s)_L$ & $(t,b)_L$\\
Position ($\mu^{-1}$) & (5.941,0) & ($-4.008$,0) & (0,0)\\\hline
Quark type           & $u_R$ & $c_R$ & $t_R$\\
Position ($\mu^{-1}$) & ($-8.347$,0) & (1.815,0) & ($-0.941$,0)\\\hline
Quark type           & $d_R$ & $s_R$ & $b_R$\\
Position ($\mu^{-1}$) & ($-8.421$,0) & (2.219,2.332) & ($-1.253$,2.767)\\
\hline
\end{tabular}
\caption{Location of left-handed and right-handed quarks inside the
thick brane in units of $\mu^{-1}$ in a model with two split-fermion
dimension\cite{11}.}
\label{tab1}
\end{table}

Next, we discuss the constraints on the parameter space of our model.
The width $\mu^{-1}$ of the wave functions should be smaller than the
brane thickness $L$ but, larger than the scale of the ultraviolet cutoff
$1/M_*$: 

\begin{equation}\label{eq2}
L^{-1} < \mu < M_*\, .
\end{equation}
A second condition is to have the four-dimensional Yukawa top coupling
$\lambda_t$ perturbative at $M_*$:
\begin{equation}\label{eq3}
\frac{M_*}{L^{-1}} <  2\pi \left( \frac{4\pi}{\lambda_t} \right)^2 \approx \frac{992.2}{{\lambda_t}^2} \, .
\end{equation}
For the field theory description to be valid through the wall $\mu^2 L <
M_*$. 
Combined with Eq.~(\ref{eq3}) this gives

\begin{equation}\label{eq4}
L_f<\frac{\mu}{L^{-1}} < \frac{31.5}{\lambda_t}\, .
\end{equation}
Here, we consider $\lambda_t<1$ or close to $1$. 
The left hand side of the condition (\ref{eq4}) restricts all the fermions to be accommodated inside the brane of width $L$, where $L_f=14.362$ is the total distance between the position of $d_R$ and $(u,d)_L$. 

In Ref.~\cite{47}, another condition is mentioned, i.e.,
$R^{-1} \sim L^{-1} > 100$~TeV.
However, this constraint comes from the flavour changing neutral
currents mediated by Kaluza-Klein gauge bosons in a compactified
five-dimensional  Standard Model, where the fifth dimension $x_5$ has
been compactified on the orbifold $S^1/Z_2$, a circle of radius $R$ with
the identification $x_5 \to -x_5$.
This condition disfavours the scenario of
TeV-strings~\cite{49} and will not be followed in this paper.

\subsection{String balls versus black holes}

In order to be able to neglect the quantum gravity effects which
strongly modify the classical general relativistic black hole solutions,
one must consider black holes of large number of states.
We will assume the commonly used condition that, the entropy of the black
hole must be greater than 25, or its mass $m > 5M_D$ ~\cite{50}. We consider the mass $M_\mathrm{min}^\mathrm{c}=5 M_D$, to be the minimum
classical mass of the black hole, i.e., the minimum mass at which the
classical general relativistic description is still valid. In this case, the window for the production of  classical black hole is small.

If one wishes to consider the LHC results on the minimum limits on the Planck scale to date~\cite{51}, high mass black holes ($m > 5M_D$) cannot be produced in three and four extra dimensions for $\sqrt{s}=(8$--14) TeV . This is because the black hole cross-section drops with increasing black
hole mass, due to the steeply falling parton density distributions in 
the protons with increasing parton centre-of-mass energy. The experimental limits on the fundamental Planck scale in \cite{51} are set using only the direct graviton emission searches from the LHC experiments. However, for large number of extra dimensions the limits are still not so stringent. Other direct limits come from non-observation of black hole events come from the simplest models (e.g. no split-splitting of fermions), which we believe cannot be taken at face value (in order to make reliable predictions one has to work with phenomenologically valid models without fast proton decay, large neutron-antineutron mixing, large mixing between leptons etc). Therefore, every number in the Table 1 in \cite{51} has to be taken in the proper context. We do not consider these limits in our paper. In the case that, the fundamental Planck scale is greater than the current experimental limits, one has to apply a proper vertical cut in our figures at the corresponding value of $M_D$.

The string entropy in any number of dimensions is proportional to the mass of string 

\begin{equation}\label{eq5}
S \sim \frac{m}{M_\mathrm{s}}\, ,
\end{equation}
\begin{center}
\begin{figure}[t]
\centering
\hspace{+1.2cm}\includegraphics[width=10cm]{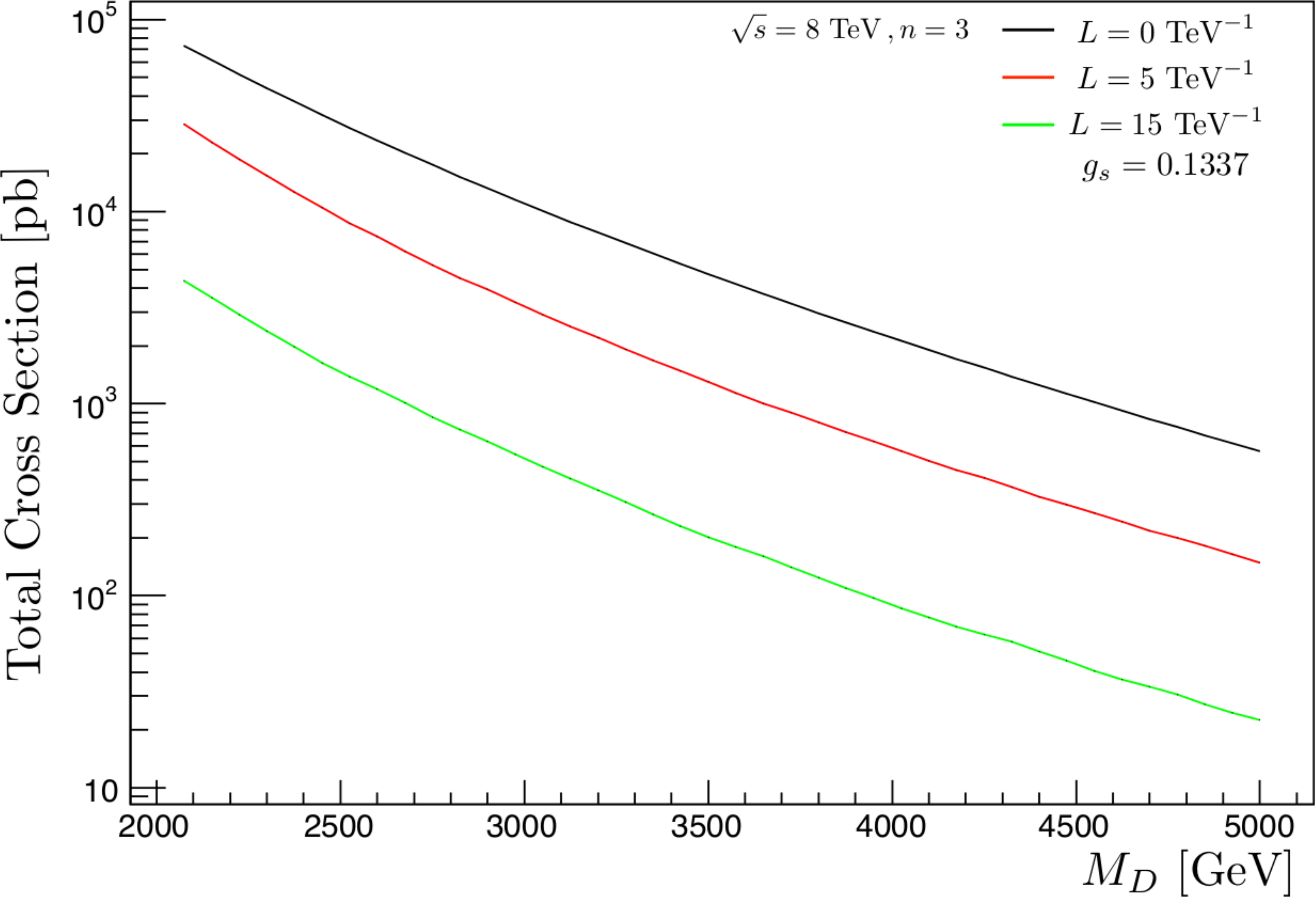}\newline\newline  \includegraphics[width=10cm]{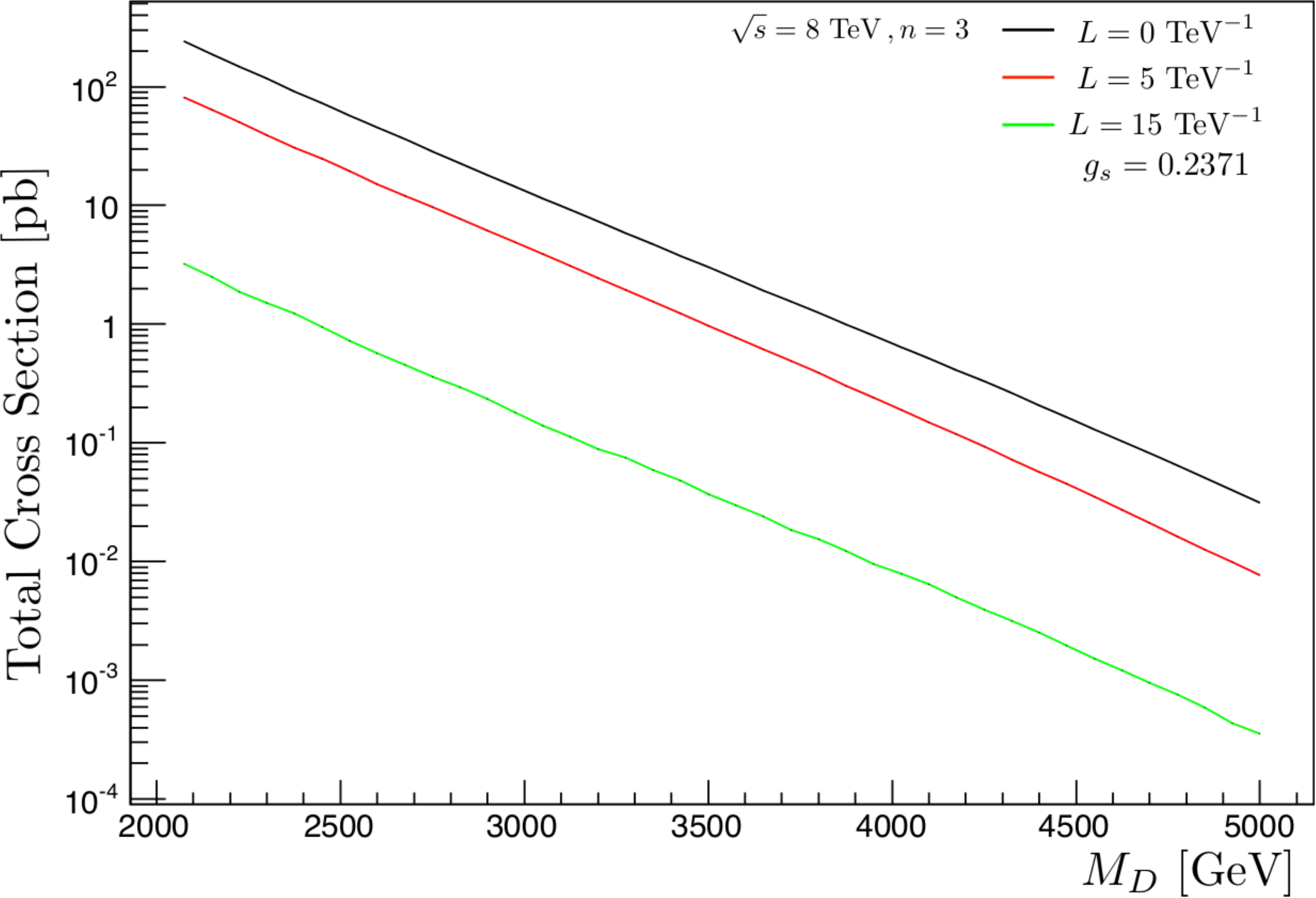}
   \caption{ \label{Fig_1}   The first plot shows total cross-section of string ball production for $\sqrt{s}=8$ TeV, $n=3$ and L=0, 5, 15 $\text{TeV}^{-1}$ and string coupling constant $g_s=0.1337$. The second plot shows total cross-section of string ball production for $\sqrt{s}=8$ TeV, $n=3$ and L=0, 5, 15 $\text{TeV}^{-1}$ and string coupling constant $g_s=0.2371$. \label{Fig_1}}
\end{figure}
\end{center} 
where $M_\mathrm{s}$ is the string scale related to the fundamental
Planck scale $M_D$.
For example, if we consider a model with $n$ large extra dimensions ($n
\le 6$) and $q$ small dimensions ($q = 6 - n$) compactified on
a $q$-torus of radius equal to the string length $l_\mathrm{s} \sim
1/M_\mathrm{s}$ then,

\begin{equation}\label{eq6}
M_\mathrm{s}  = \frac{g_\mathrm{s}^{2/(n+2)}}{(2\pi)^{(n-6)/(n+2)}} M_D\, , 
\end{equation}
where $g_\mathrm{s}$ is the dimensionless string coupling constant in
weakly-coupled string theory \cite{35}. 
For the string theory to be perturbative, $g_\mathrm{s}$ must be less
than unity. An example of a typical value of $g_s$ in theories with large extra dimensions with values of $g_s$ between 0.02 and 0.2 have been discussed in \cite{52}. There are other results \cite{13,14} which, they take into account string coupling constant $g_s$ to be close to one. Relation (\ref{eq6}) is model dependent, with respect to the specific compactification schemes and the string theory model under consideration. For example, one can find a different relation between $M_s$ and $M_D$ \cite{36}. Therefore, we write the relation (\ref{eq6}) in the following form:
\begin{equation}\label{eq7}
M_\mathrm{s} =\gamma g_\mathrm{s}^{2/(n+2)} M_D \, ,
\end{equation} where, the exact numerical coefficient $\gamma$ is model dependent.

\begin{center}
\begin{figure}[t]
\centering
 \hspace{+1.2cm}\includegraphics[width=10cm]{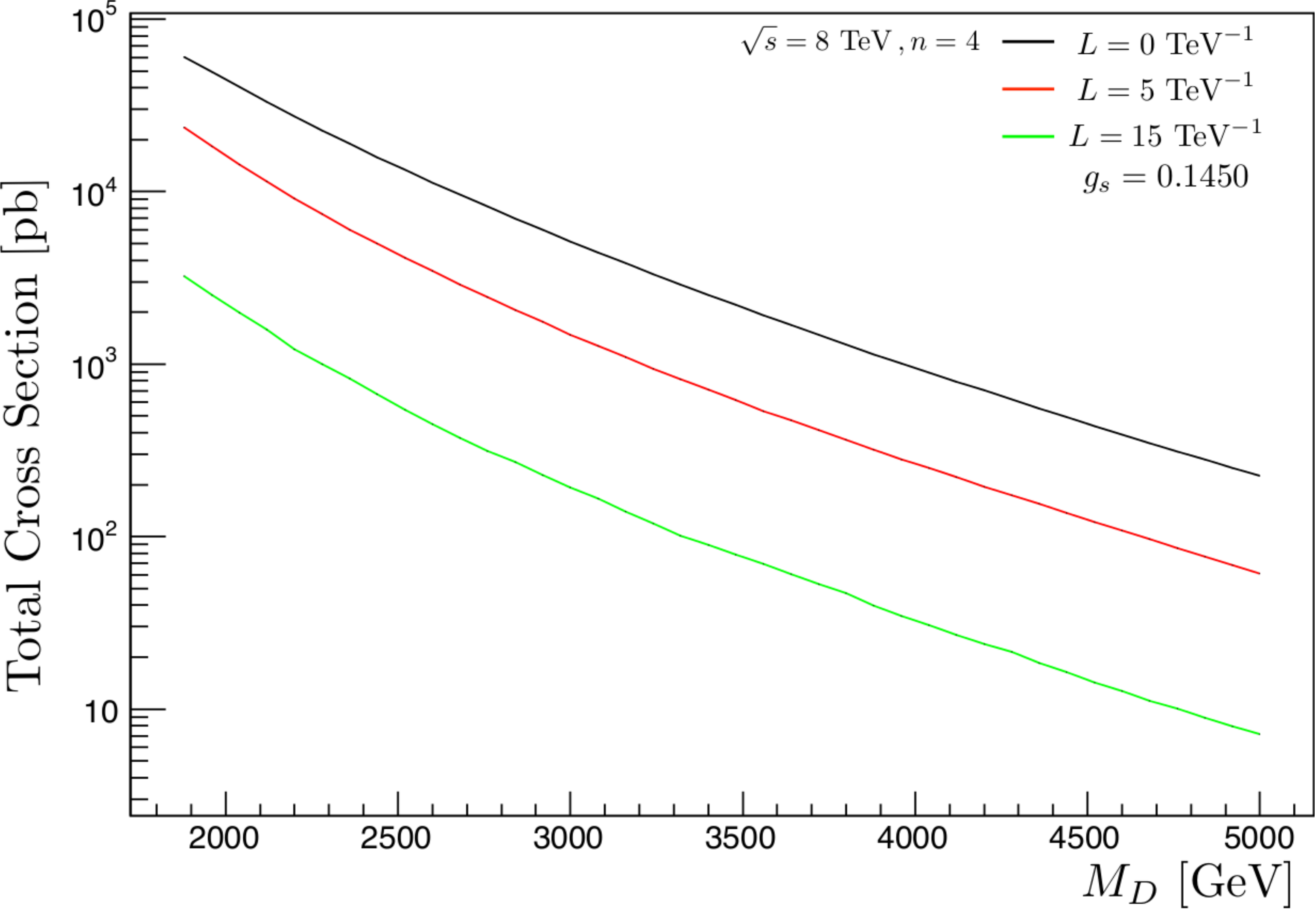}\newline\newline  \includegraphics[width=10cm]{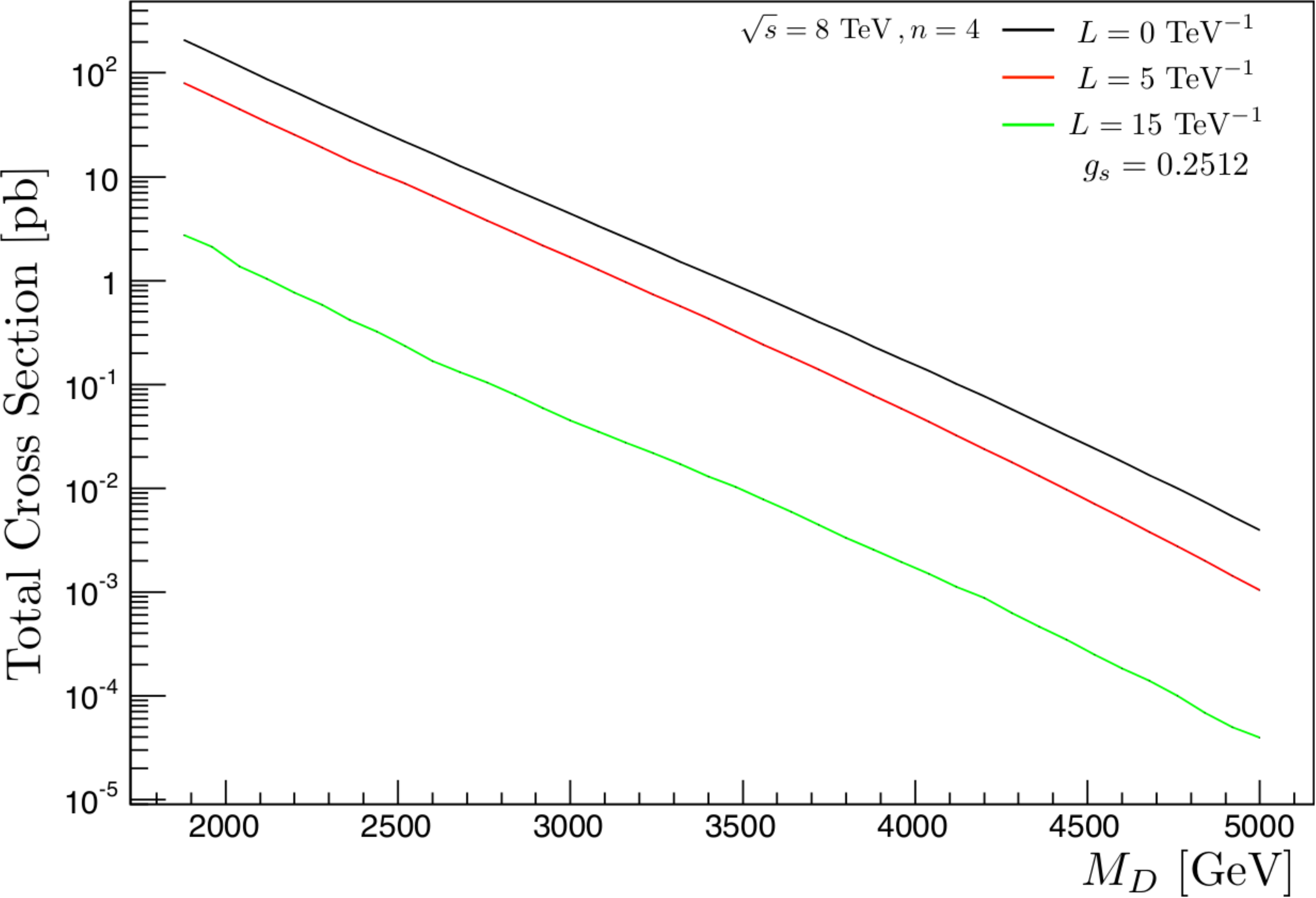}
    \caption{ \label{Fig_2}   The first plot shows total cross-section of string ball production for $\sqrt{s}=8$ TeV, $n=4$ and L=0, 5, 15 $\text{TeV}^{-1}$ and string coupling constant $g_s=0.1450$. The second plot shows total cross-section of string ball production for $\sqrt{s}=8$ TeV, $n=4$ and L=0, 5, 15 $\text{TeV}^{-1}$ and string coupling constant $g_s=0.2512$. \label{Fig_2}}
\end{figure}
\end{center}
We assume that, in a quantum theory of gravity there will be a space of states $H(E, J)$ which describes the possible quantum states of the gravitational field which have energy $E$ and angular momentum $J$. 
Assume that, there is a region in the space of states $H(E,J)$, which contains 
the quantum states which correspond to the black hole $H_\mathrm{BH}(E,J)$, 
and a region $H_\mathrm{s}(E,J)$ which, contains the corresponding excited
states of the strings, and that the logarithm of the number of states in
the string region is proportional to the relevant entropy.
Further, we assume that, the rates for transitions between the states in the
black hole and the string regions are on the order of the rates for
transitions between states within the regions, and that all of these
rates are on the order of $1/E$. 
Given this, one may expect that the relative entropy of states in the
two regions gives the relative probabilities that the system will be
found in each region, independently of the initial conditions.

Suppose that, a black hole of mass $m$ is formed in the
collider with $M_D < M_\mathrm{min}^\mathrm{c} < m$. Also consider that, the
black hole is in empty space and evaporating according to the Hawking 
radiation process.
As the energy decreases we consider the probability for the black hole
state to make a transition to one of the massive string states.
There exists a critical value of the mass where the entropies are equal,
above this value the entropy of the black hole is greater than the
entropy of the string ball, and the black hole state is the favourable
state, below this value of mass the string state is the favourable state.
Thus, it is entropically favourable for the black hole to make a
transition to one of the massive modes of the string at the
correspondence point, where the entropies are equal,
\begin{equation}\label{eq8}
m \sim \frac{M_\mathrm{s}}{g_\mathrm{s}^2}\, .
\end{equation}
We call this minimum mass ($M_s/g_s^2$) the stringy minimum mass of the black hole $M_{min}^s$.
The black hole leaves no remnant, i.e., the black hole can evaporate completely, avoiding the
singularity that is inevitable in the semiclassical picture \cite{31}.

Going back to our first assumption that, in a quantum theory of gravity
there will be a space of states $H(E,J)$ which describes the possible quantum
states of the gravitational field, we can view black holes and string
balls as describing different forms of the same gravitational object at
different energies. 
Thus, we expect that for centre-of-mass energies less than
$M_\mathrm{min}^\mathrm{s}$ string balls to be produced at the LHC,
rather than black holes. Namely, the correspondence between the black hole and string ball also suggests that, string balls production is dual to the black hole production, and the production cross-section for string balls will match the black hole cross-section at the correspondence point \cite{14}. In this paper, we equate the stringy minimum mass of black hole $M_{min}^s$ to the classical minimum mass of black hole $M_{min}^c$. One may still hope to copiously produce black holes lighter than $m > 5M_D$, and that, the quantum corrections will not be disastrous. For this reason, in one of our figures we consider the cross-section for string ball and black hole production in the case that $M_\mathrm{min}^\mathrm{c}=M_{min}^s=3 M_D$. 

The cross-section and event characteristics for black holes and string balls in higher dimensional models with no split-fermions have already been considered~\cite{38}. In \cite{38}, $M_\mathrm{min}^\mathrm{c}$ is taken to be $5M_D$ which is supported by their thermodynamic and Compton wavelength argument. The cross-section for the production of black holes and their angular momentum distribution in a one dimensional split-fermion model have been studied~\cite{53}.It was shown that, for a given value of the Planck scale, the total production cross-section is reduced as compared to the model with non-split-fermions. Here, we have considered string balls and black holes into a two-dimensional split-fermion model for the first time. To calculate the cross-section of the black hole or string ball production we use the equation (21) of \cite{53} which is implemented in BlackMax code. 

\section{Numerical results}\label{sec3}
The production of black holes and string balls in particle collisions
depend on the following parameters:

\begin{enumerate}
\item The parameters of the higher-dimensional space: number of
      extra dimensions $n$ and the $D$-dimensional fundamental Planck
      scale $M_D$. We use the PDG definition of the Planck scale~\cite{54}. 
      
\item The parameters of the split-fermion model: the number of
      split-fermion dimensions, (we consider a two-dimensional
      split-fermion model), the width of the fermion wave functions, the width of the split-fermion brane $L$, and the location of the fermions in the split-fermion brane taken from
      Table~\ref{tab1}.  
\item The parameters of the particle collisions: the type and
      centre-of-mass energy of the colliding partons.
      We consider proton-proton collisions at the LHC energies of 8~TeV,
     and 14~TeV. This allows us to compare our results with the current and the future run of the LHC. We use the CTEQ6.1M parton distribution
      functions~\cite{55}.     
 \item The mass, momentum, and angular momentum loss factors in the
      inelastic collision before the formation of the black hole.
    
     \item The string ball parameters: the string scale
      $M_\mathrm{s}$, the string coupling constant $g_\mathrm{s}$, and the
      minimum mass of the string ball $M_\mathrm{ms}$, which
      we take to be $3M_\mathrm{s}$.
      \end{enumerate}
\begin{center}
\begin{figure}[t]
\centering
  \hspace{+1.2cm} \includegraphics[width=10cm]{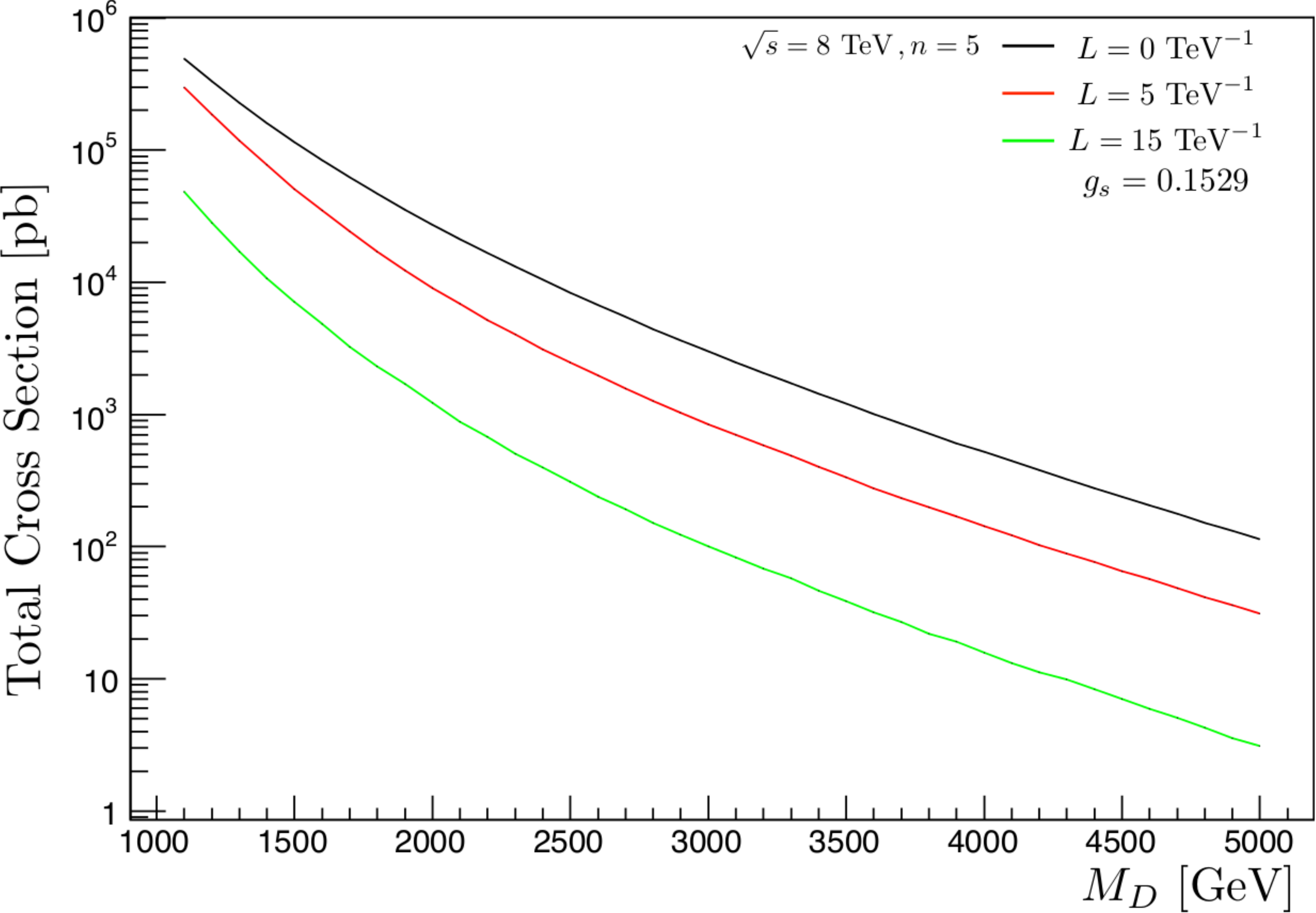}\newline\newline
    \includegraphics[width=10cm]{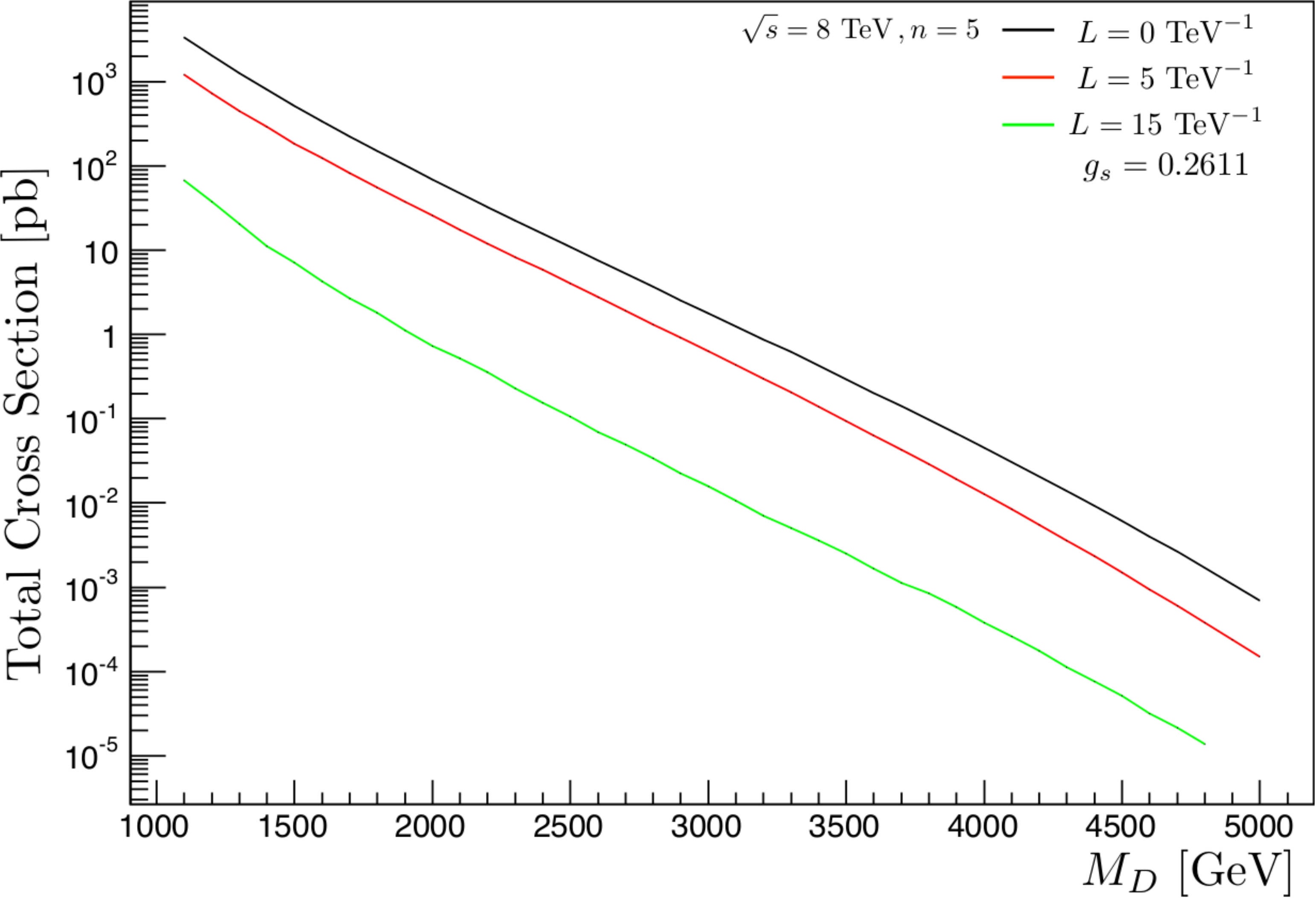}
   \caption{ \label{Fig_3}   The first plot shows total cross-section of string ball production for $\sqrt{s}=8$ TeV, $n=5$ and L=0, 5, 15 $\text{TeV}^{-1}$ and string coupling constant $g_s=0.1529$. The second plot shows total cross-section of string ball production for $\sqrt{s}=8$ TeV, $n=5$ and L=0, 5, 15 $\text{TeV}^{-1}$ and string coupling constant $g_s=0.2611$. \label{Fig_3}}
\end{figure}
\end{center}
Typically, the scales of the problem have the following hierarchy 
\begin{equation}\label{eq9}
M_\mathrm{s} < M_\mathrm{ms} < M_D < \frac{M_\mathrm{s}}{g_\mathrm{s}^2}\, .   
\end{equation}
Other parameters of the simulation are the minimum classical mass of the black hole $M_{min}^c$ and the stringy minimum mass of black hole $M_{min}^s$. It is natural to equate the stringy minimum mass of black hole $M_{min}^s$ to the classical minimum mass of black hole $M_{min}^c$.
In other words,
\begin{equation}\label{eq10}
M_\mathrm{min}^\mathrm{s}\sim\frac{M_\mathrm{s}}{g_\mathrm{s}^2} = M_\mathrm{min}^\mathrm{c} = 5 M_D\, .
\end{equation} By this condition and Eq.~(\ref{eq7}), the string scale $M_s$ and the string
coupling constant $g_s$ are fixed, if we know $\gamma$. 

In what follows we investigate the dependence of the cross-section production of string balls and black holes on the above parameters. 

\begin{figure}[tbp]
\hfill
  \includegraphics[width=10cm]{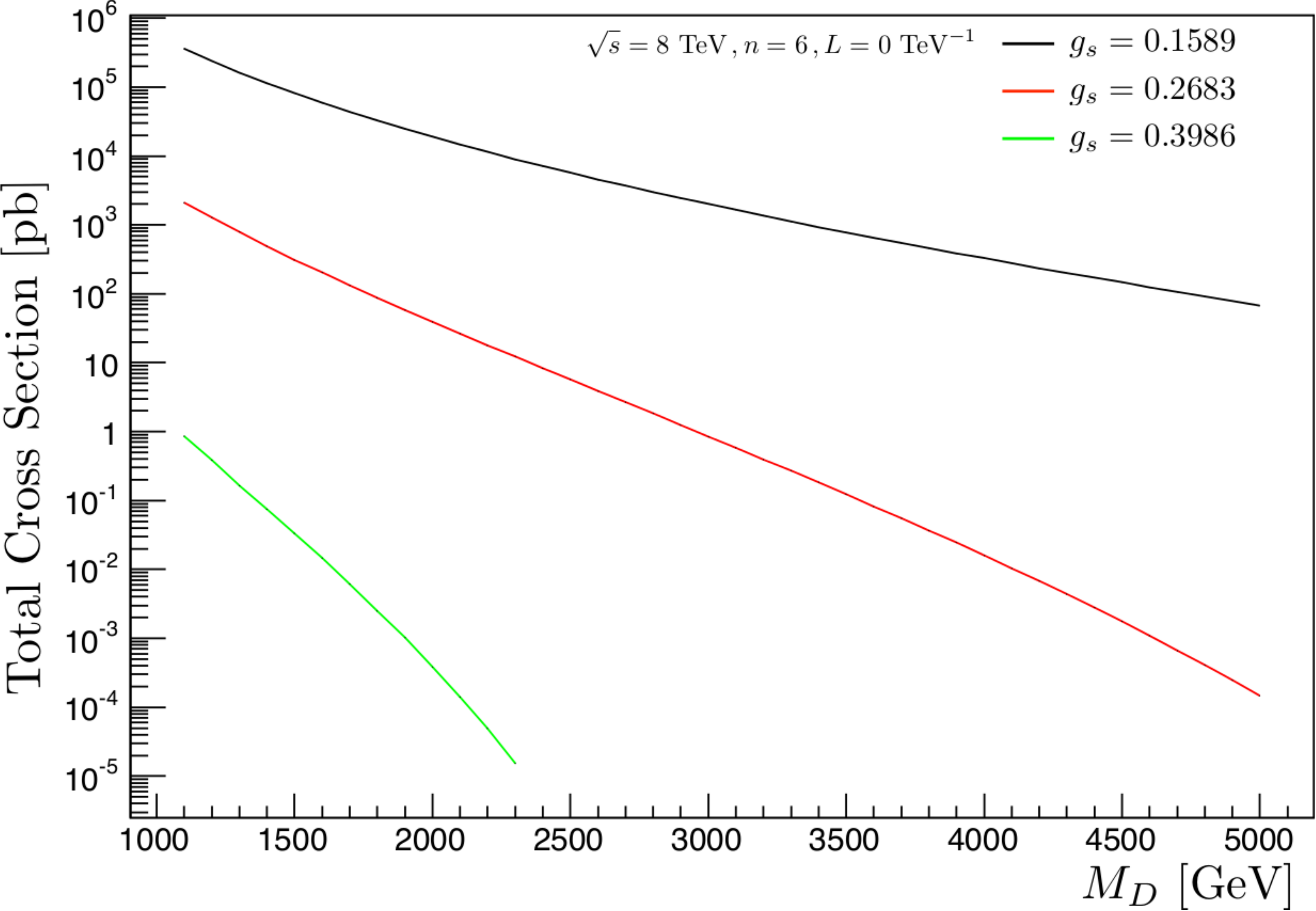}
  \hfill  \, { }
  \caption{ \label{Fig_4}  Total cross-section of string ball production for $\sqrt{s}=8$ TeV, $n=6$ and L=0 $\text{TeV}^{-1}$ and different values of string coupling constant ($g_s$). 
   \label{Fig_4}}
\end{figure}

\begin{figure}[tbp]
\hfill
  \includegraphics[width=10cm]{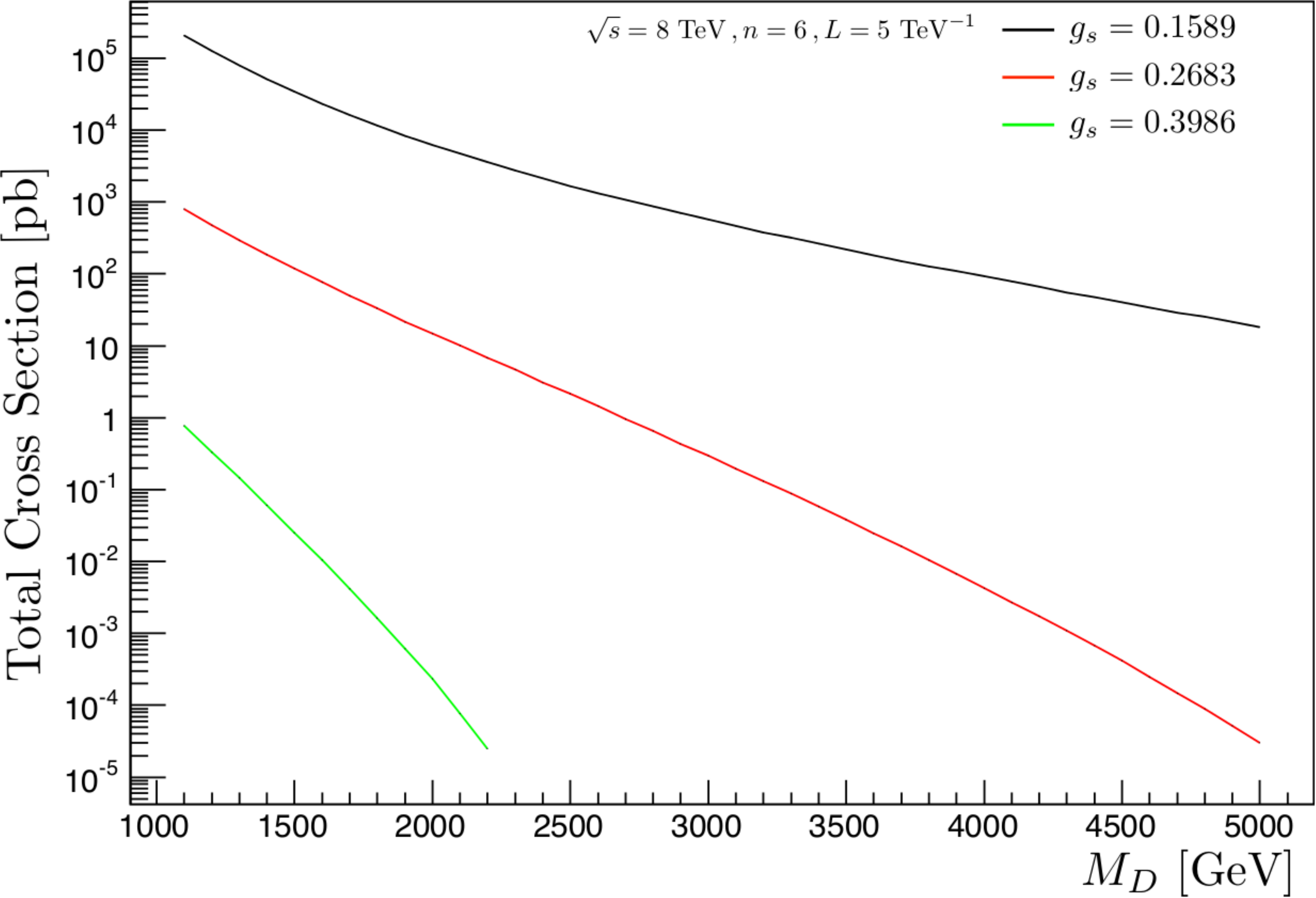}
  \hfill  \, { }
  \caption{ \label{Fig_5}  Total cross-section of string ball production for $\sqrt{s}=8$ TeV, $n=6$ and L=5 $\text{TeV}^{-1}$ and different values of string coupling constant ($g_s$). 
   \label{Fig_5}}
\end{figure}
\begin{figure}[tbp]
\hfill
  \includegraphics[width=10cm]{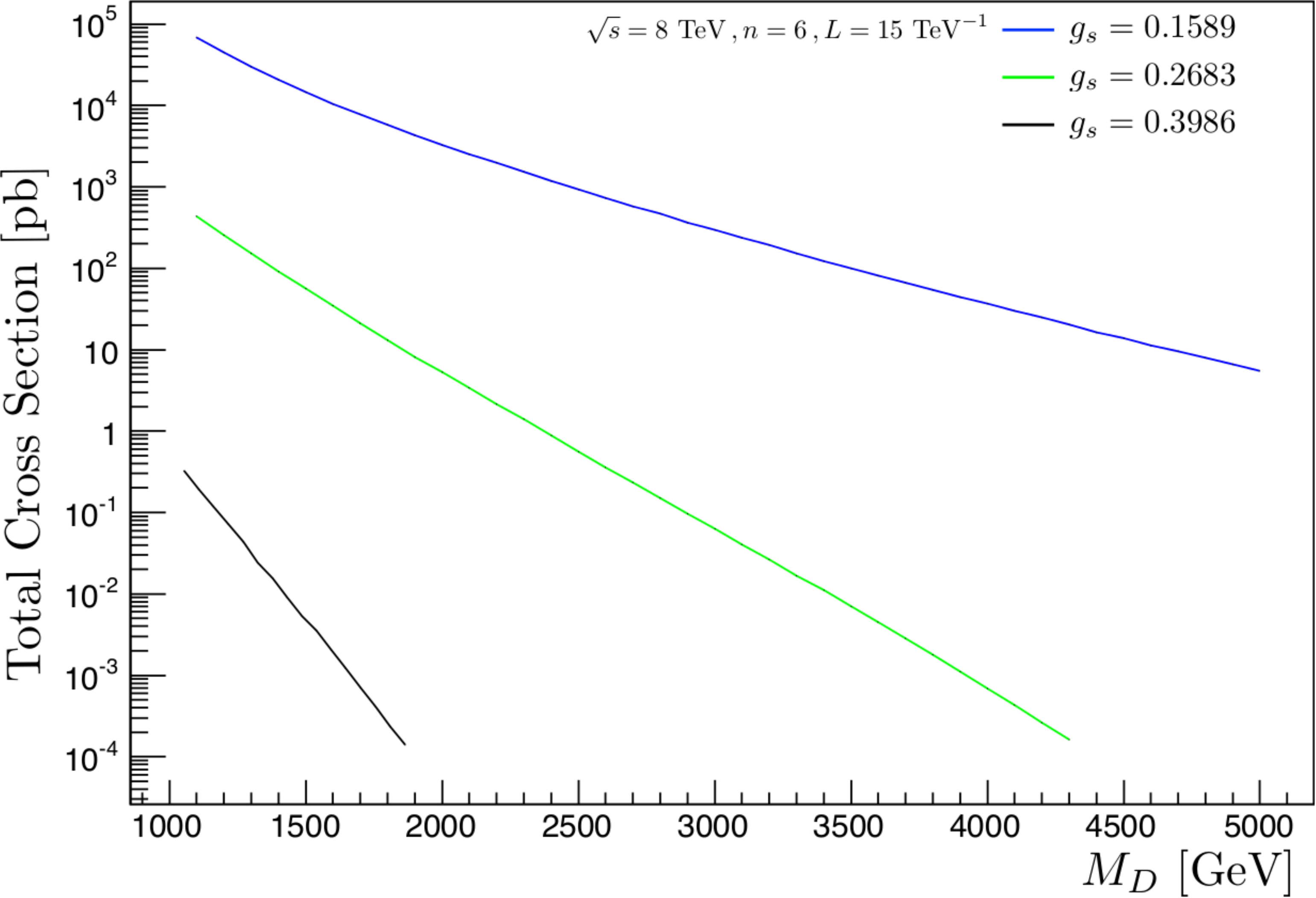}
  \hfill  \, { }
  \caption{ \label{Fig_6}  Total cross-section of string ball production for $\sqrt{s}=8$ TeV, $n=6$ and L=15 $\text{TeV}^{-1}$ and different values of string coupling constant ($g_s$). 
   \label{Fig_6}}
\end{figure}
\begin{figure}[tbp]
\hfill
 \includegraphics[width=10cm]{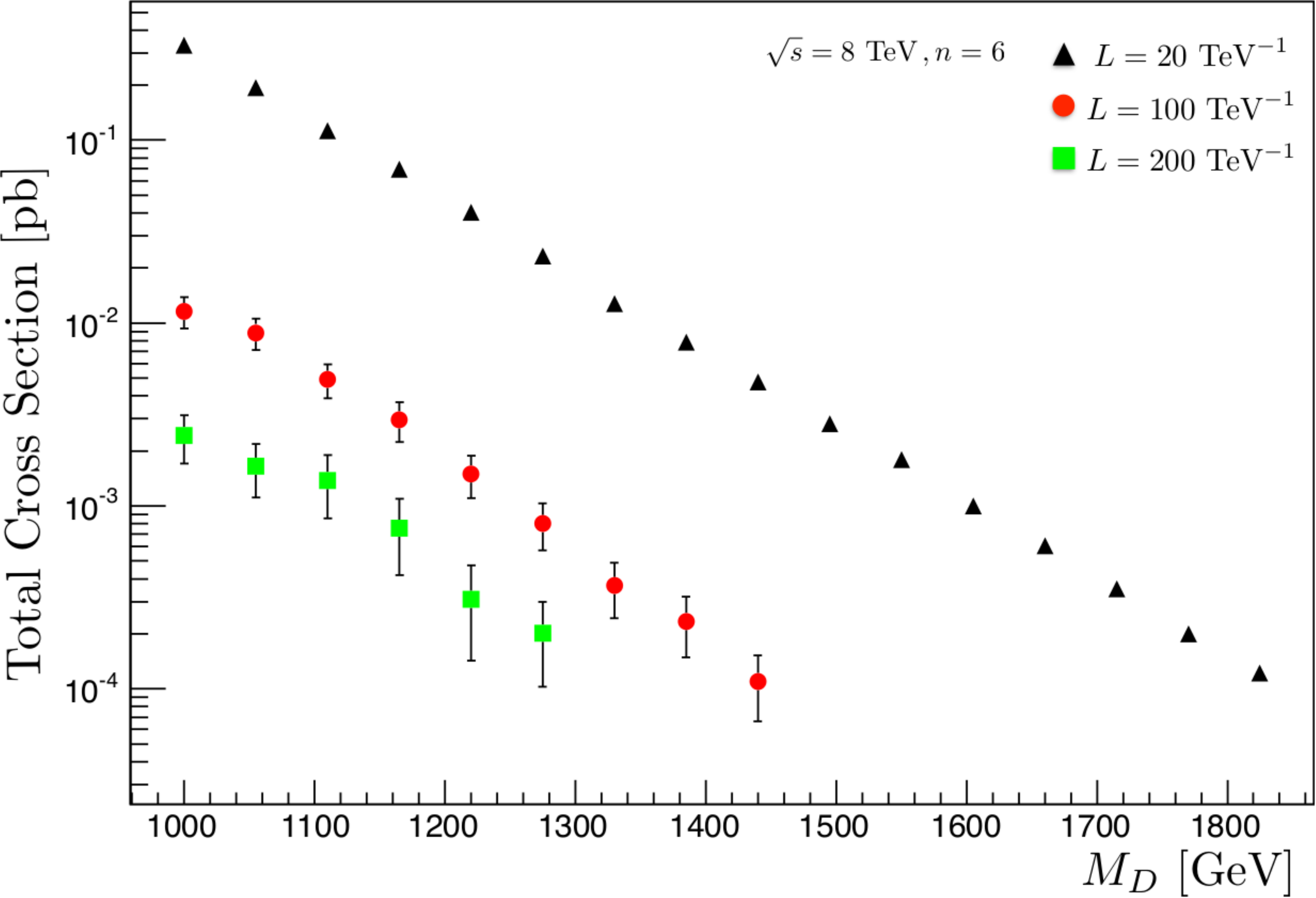}
  \hfill  \, { }
  \caption{ \label{Fig_7} Total cross-section of string ball production for n=6, $\sqrt{s}=8$ TeV, and for different values of $L$.
   \label{Fig_7}}
\end{figure}

\vspace{-8cm}
\begin{center}
\begin{figure}[t]
\centering
 \hspace{+1.2cm}\includegraphics[width=10cm]{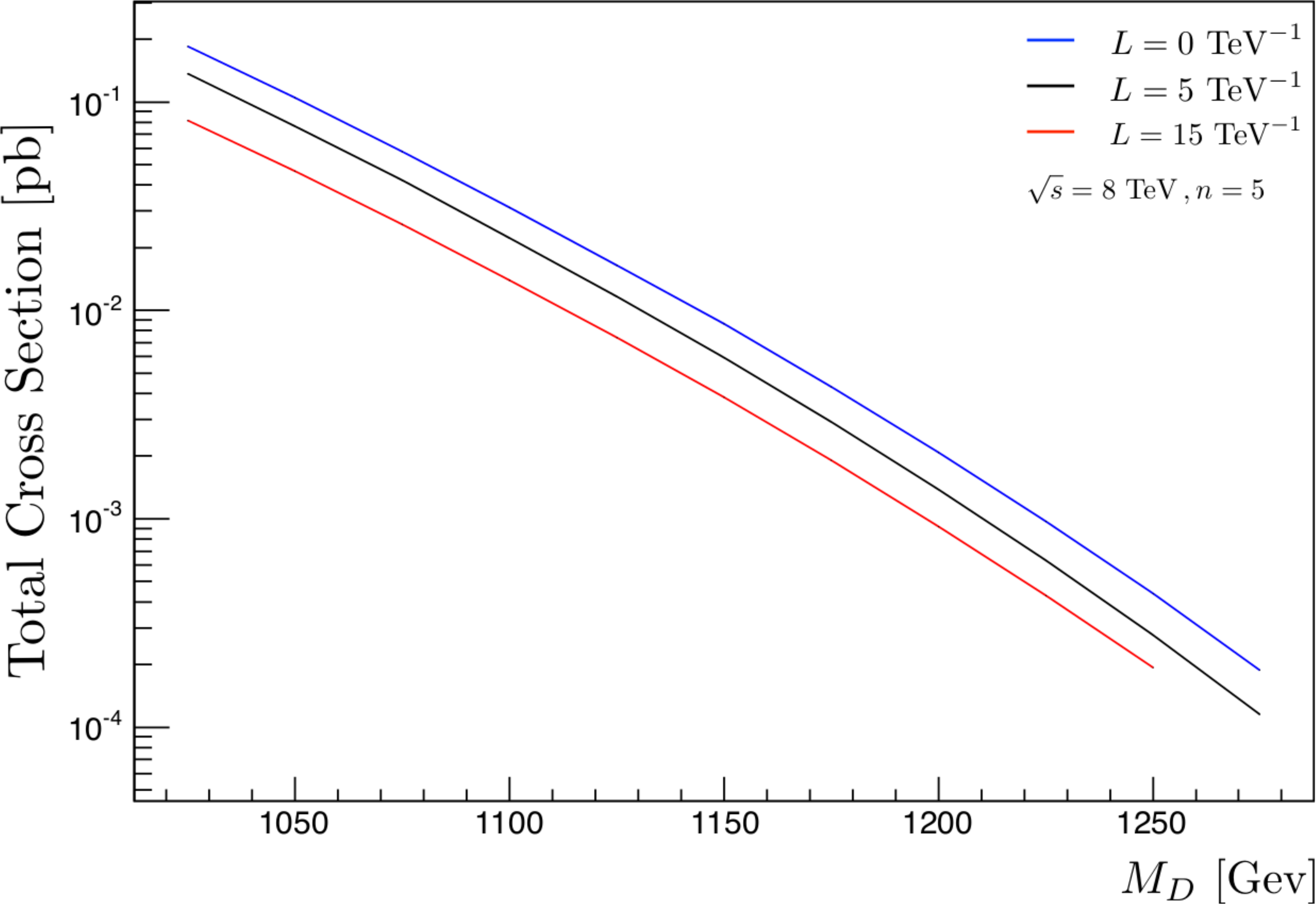}\newline\newline  \includegraphics[width=10cm]{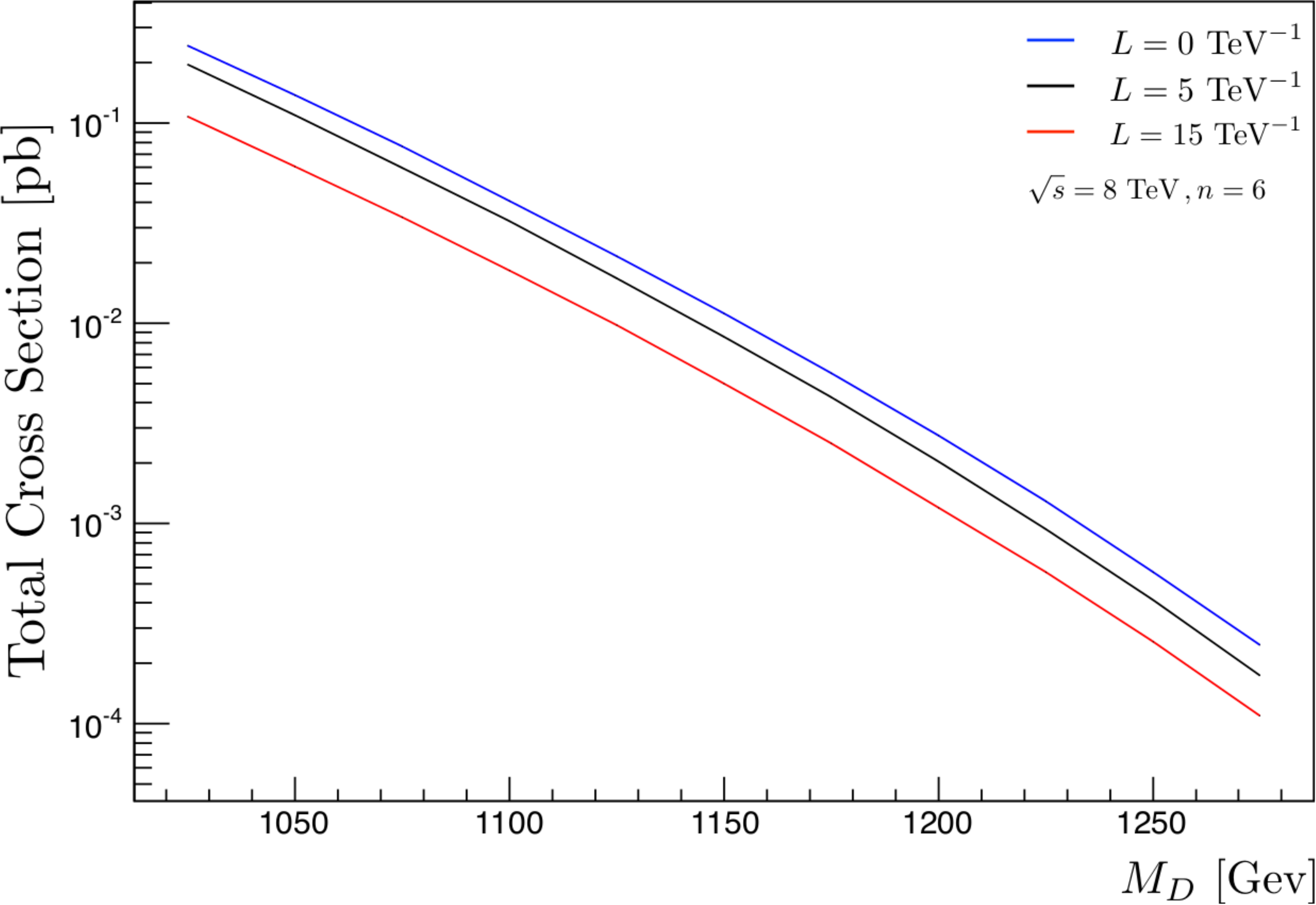}
   \caption{ \label{Fig_8}   The first plot shows total cross-section of black hole production for $\sqrt{s}=8$ TeV, $n=5$ and L=0, 5, 15 $\text{TeV}^{-1}$. The second plot shows total cross-section of black hole production for $\sqrt{s}=8$ TeV, $n=6$ and L=0, 5, 15 $\text{TeV}^{-1}$. \label{Fig_8}}
\end{figure}
\end{center}
\begin{center}
\begin{figure}[htb]
\centering
\hspace{+1.2cm}\includegraphics[width=10cm]{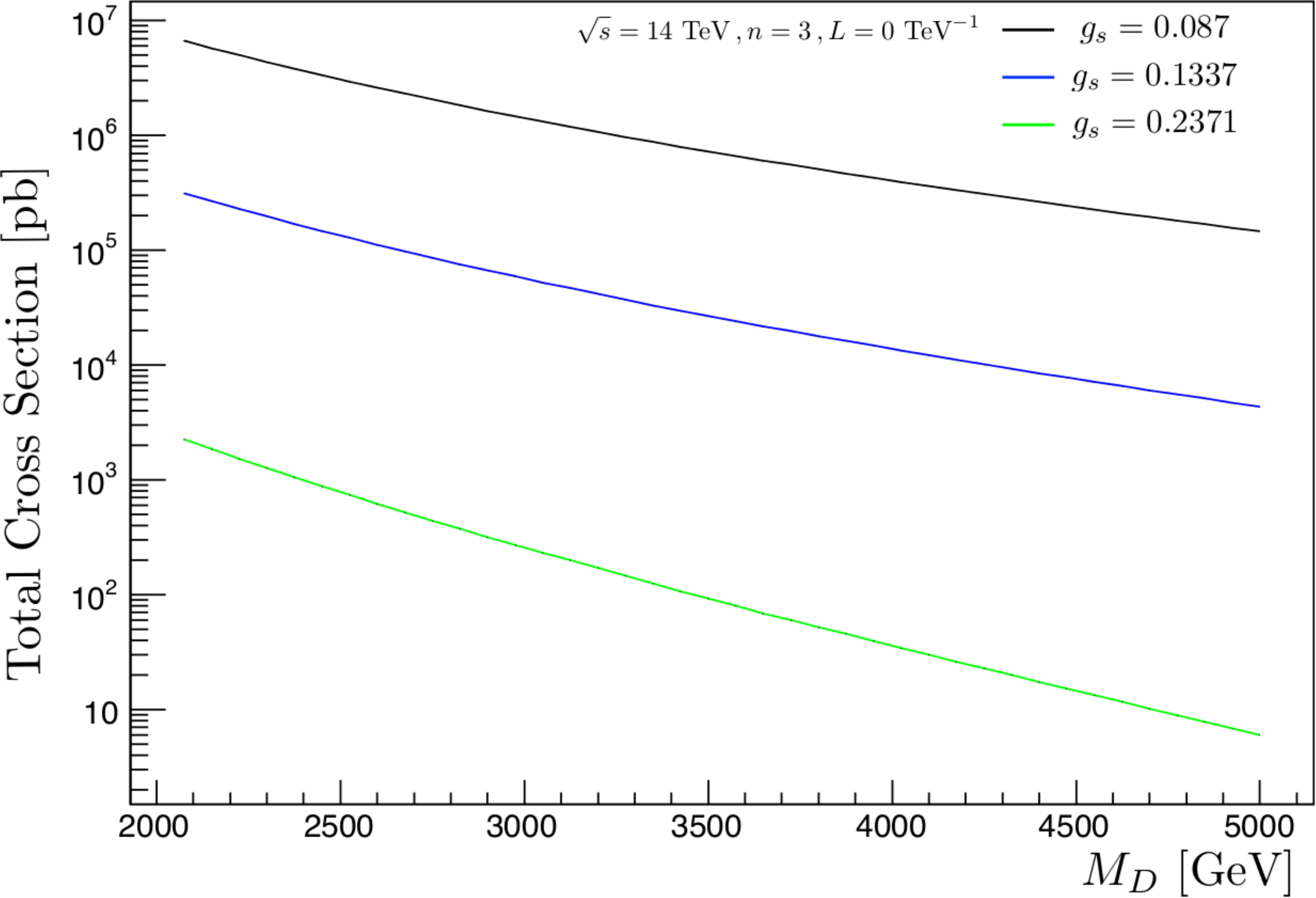}\newline\newline  \includegraphics[width=10cm]{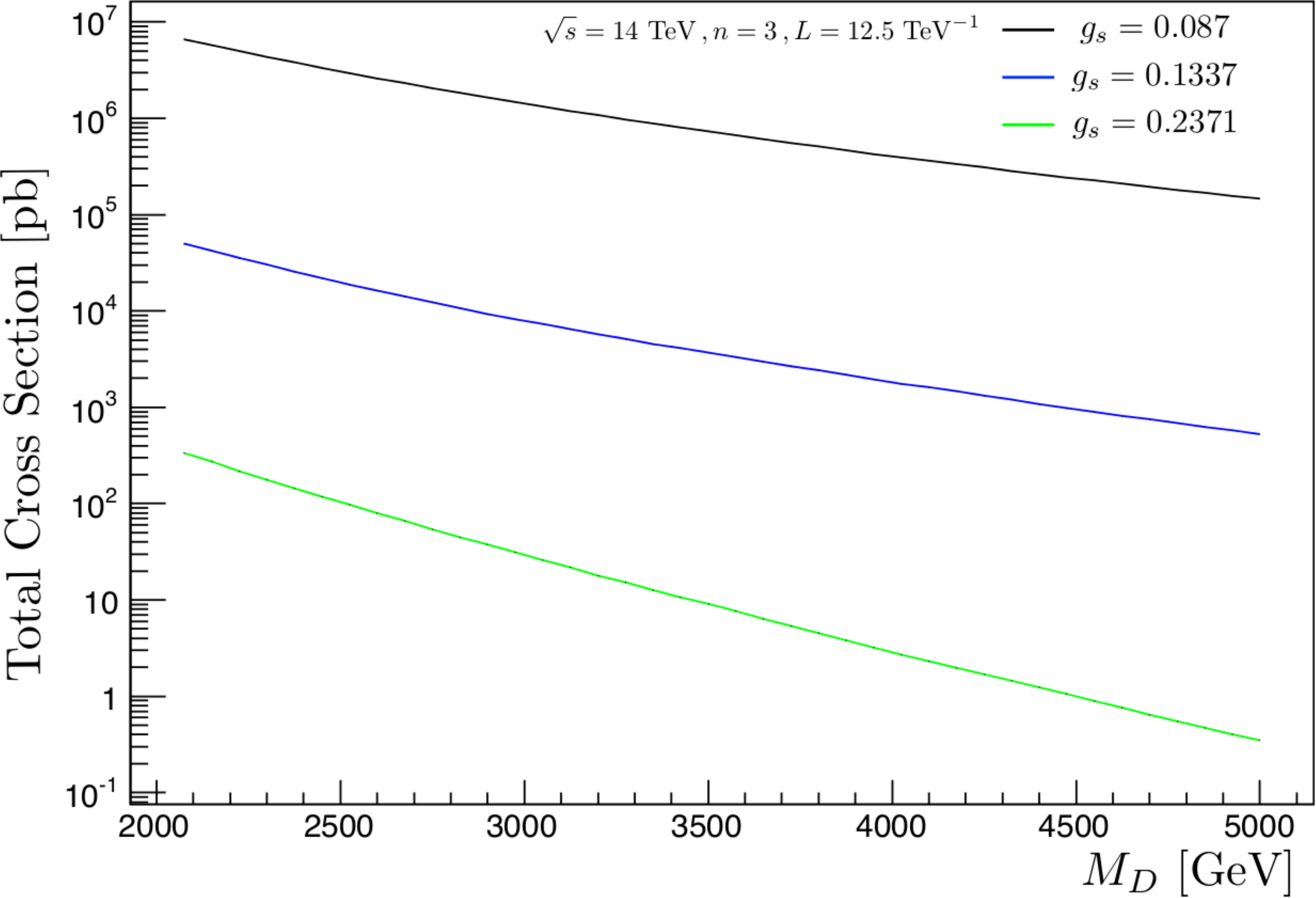}
    \caption{ \label{Fig_9}   The first plot shows the total cross-section of string ball production for $\sqrt{s}=14$ TeV, $n=3$ and L=0 $\text{TeV}^{-1}$ and different values of string coupling constant ($g_s$). The second plot shows the total cross-section of string ball production for $\sqrt{s}=14$ TeV, $n=3$ and L=12.5 $\text{TeV}^{-1}$ and different values of string coupling constant ($g_s$). \label{Fig_9}}
\end{figure}
\end{center}
\begin{center}
\begin{figure}[htb]
\centering
\hspace{+1.2cm} \includegraphics[width=10cm]{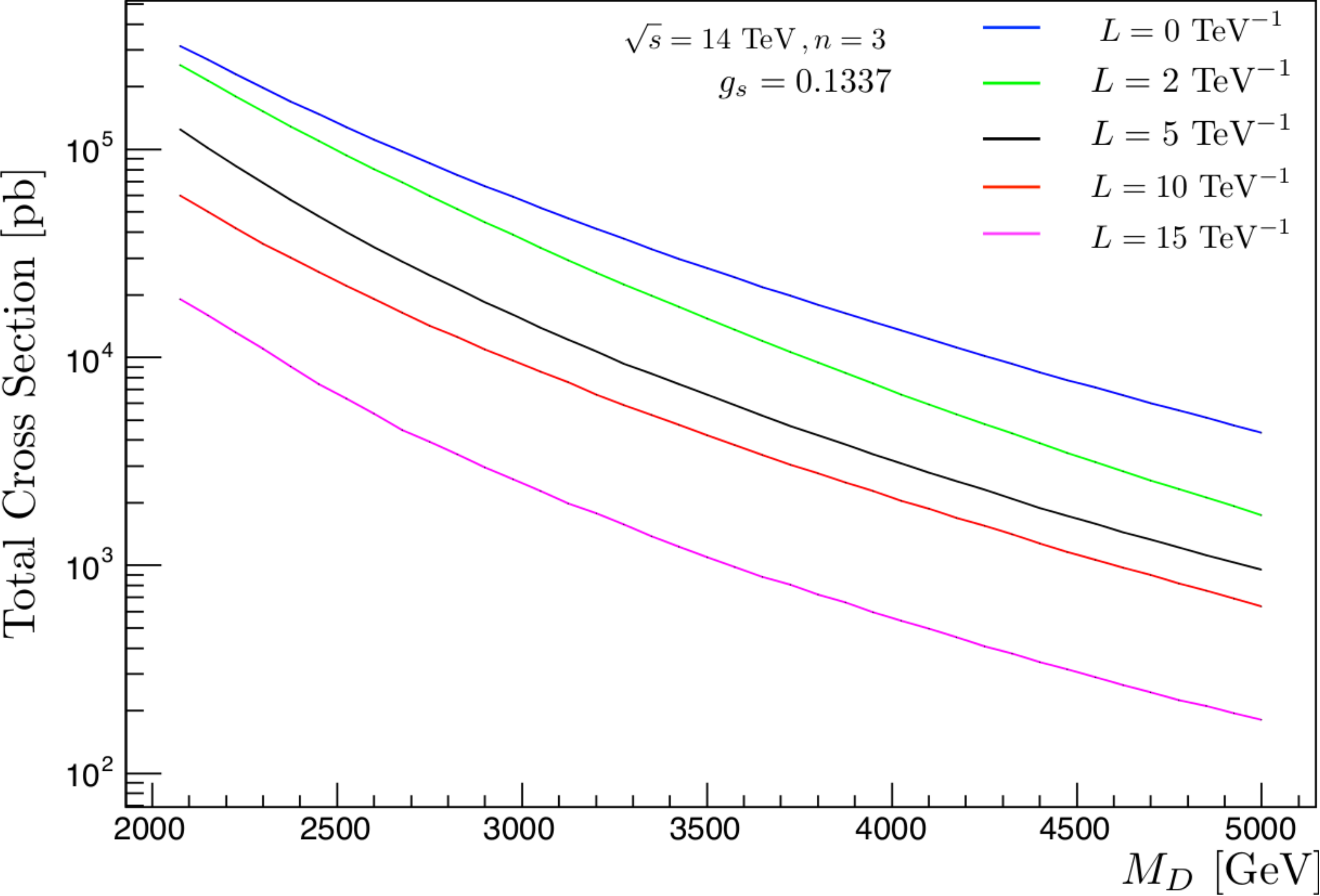}\newline\newline    \includegraphics[width=10cm]{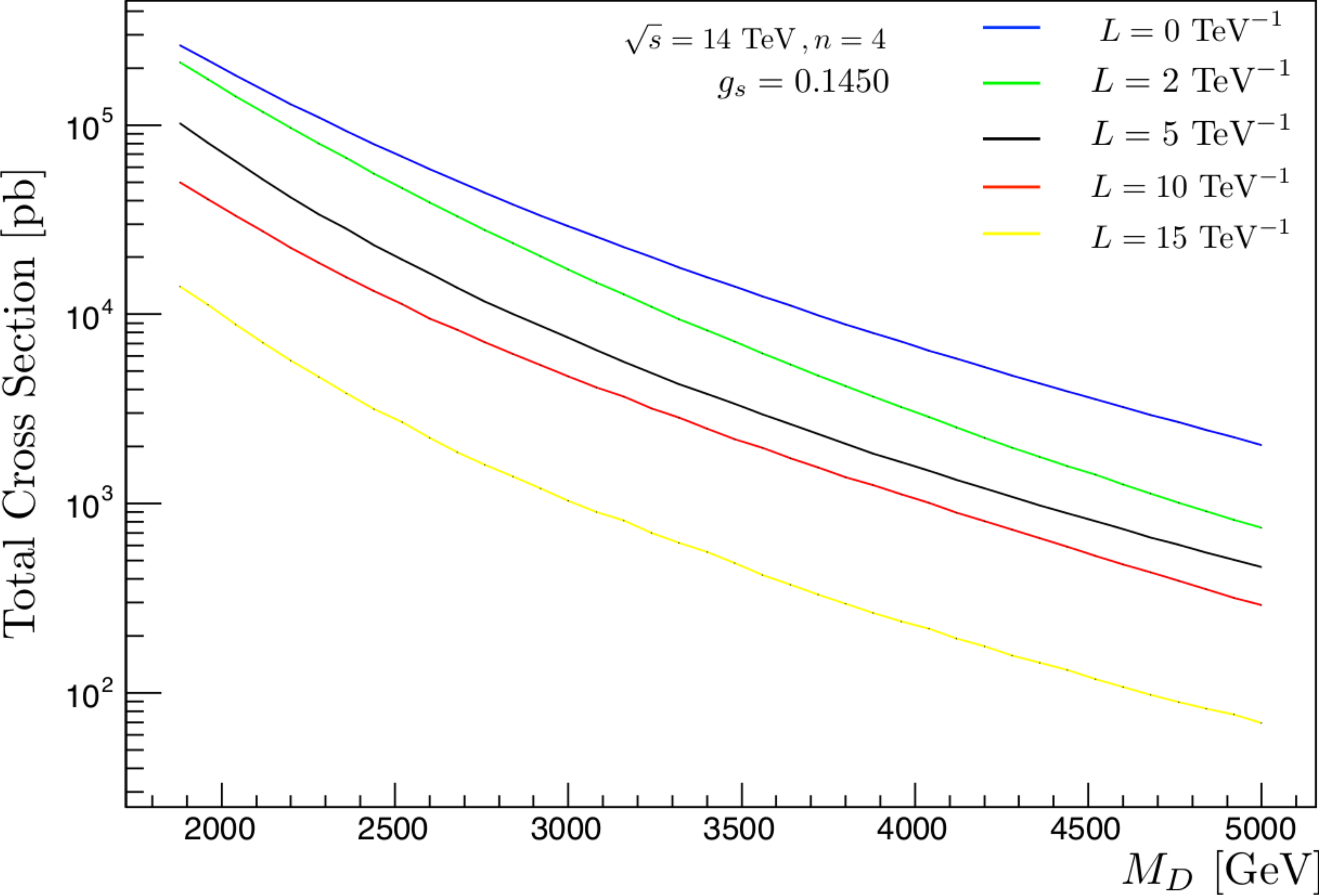}
  \caption{ \label{Fig_10}   The first plot shows the total cross-section of string ball production for $\sqrt{s}=14$ TeV, $n=3$ and L=0, 2, 5, 10, 15 $\text{TeV}^{-1}$. The second plot shows the total cross-section of string ball production for $\sqrt{s}=14$ TeV, $n=4$ and L=0, 2, 5, 10, 15 $\text{TeV}^{-1}$. \label{Fig_10}}
\end{figure}
\end{center}
\begin{center}
\begin{figure}[htb]
\centering
\hspace{+1.2cm} \includegraphics[width=10cm]{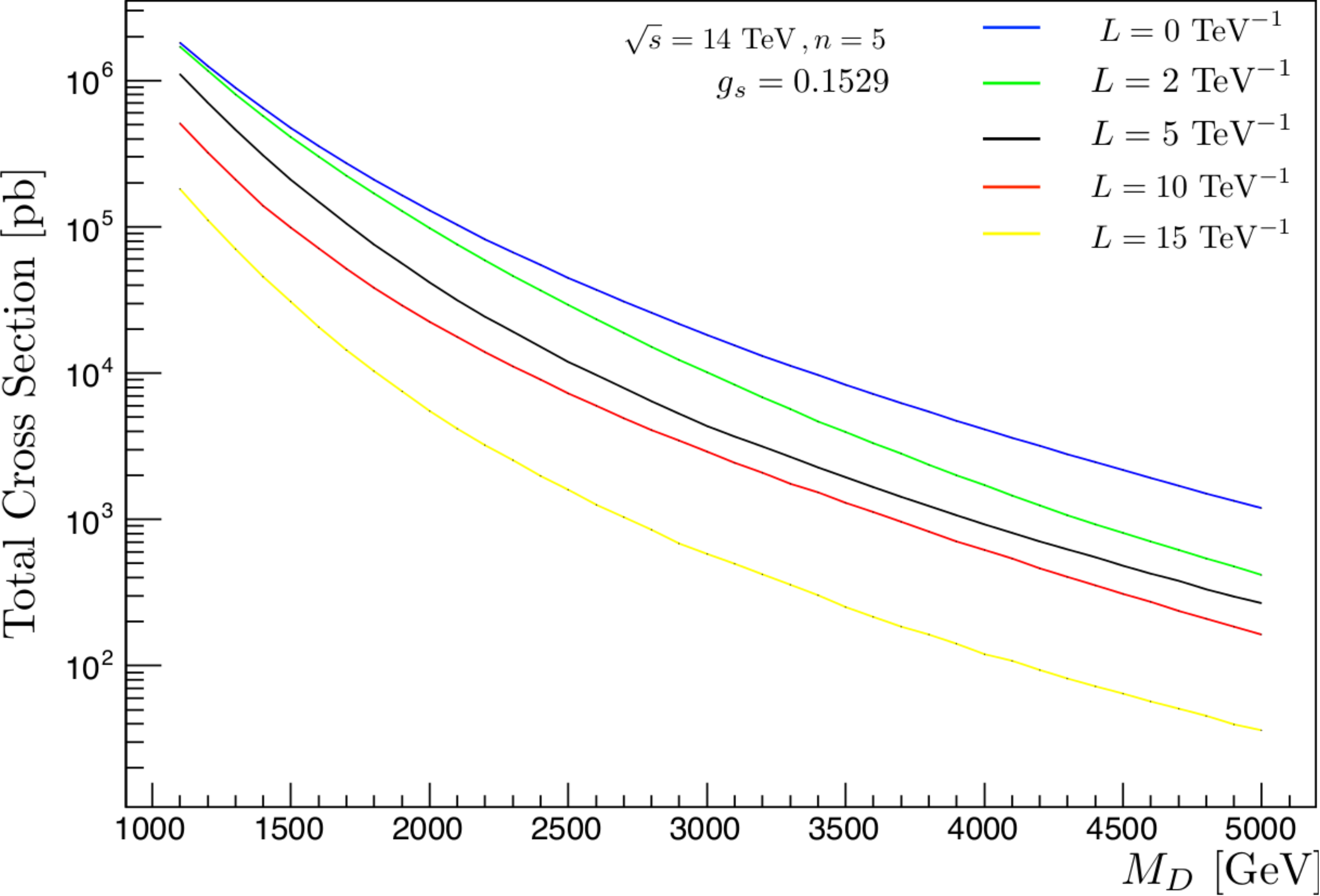}\newline\newline  \includegraphics[width=10cm]{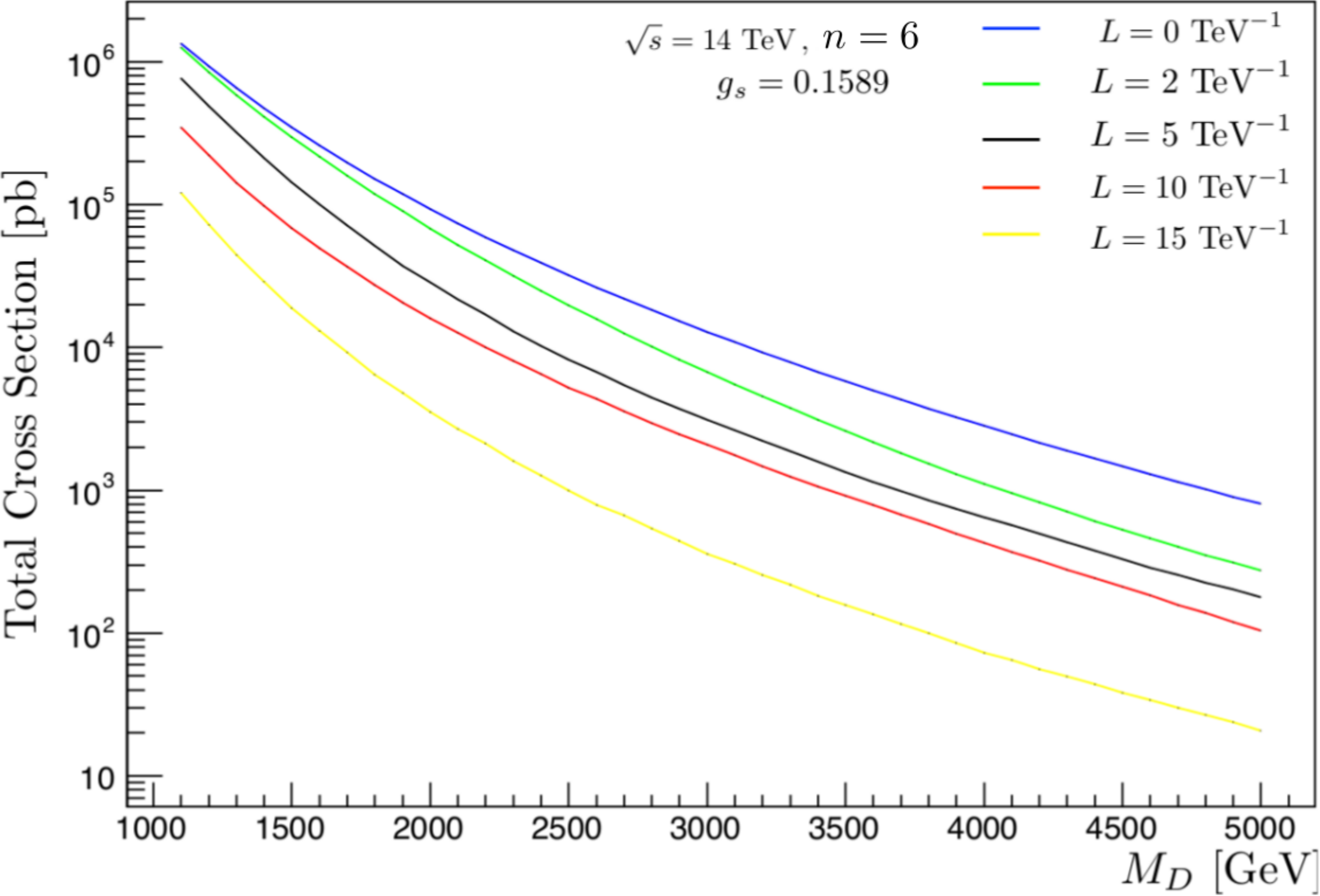}
    \caption{ \label{Fig_11}   The first plot shows the total cross-section of string ball production for $\sqrt{s}=14$ TeV, $n=5$ and L=0, 2, 5, 10, 15 $\text{TeV}^{-1}$. The second plot shows the total cross-section of string ball production for $\sqrt{s}=14$ TeV, $n=6$ and L=0, 2, 5, 10, 15 $\text{TeV}^{-1}$. \label{Fig_11}}
\end{figure}
\end{center}
\begin{figure}[tbp]
\hfill
  \includegraphics[width=10cm]{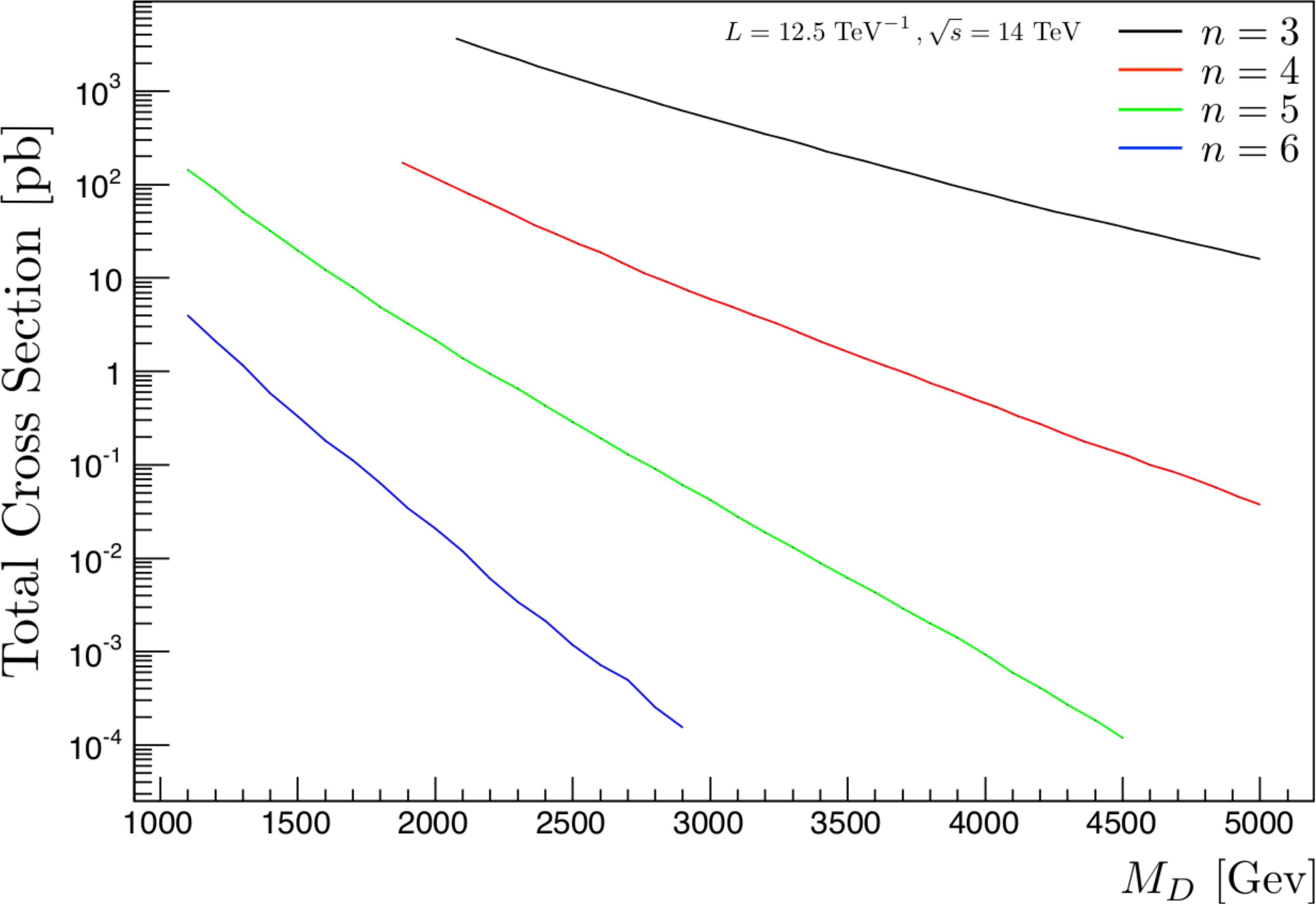}
  \hfill  \, { }
  \caption{ \label{Fig_12}  Total cross-section of string ball production for L=12.5 $\text{TeV}^{-1}$, $\sqrt{s}=14$ TeV, for different number of extra dimensions while parameter $\gamma$ varies with respect to dimensions. From the top the values of $g_s$ are the following: 0.1836, 0.2636, 0.3355, 0.3986.
   \label{Fig_12}}
\end{figure}

\begin{center}
\begin{figure}[t]
\centering
 \hspace{+1.2cm} \includegraphics[width=10cm]{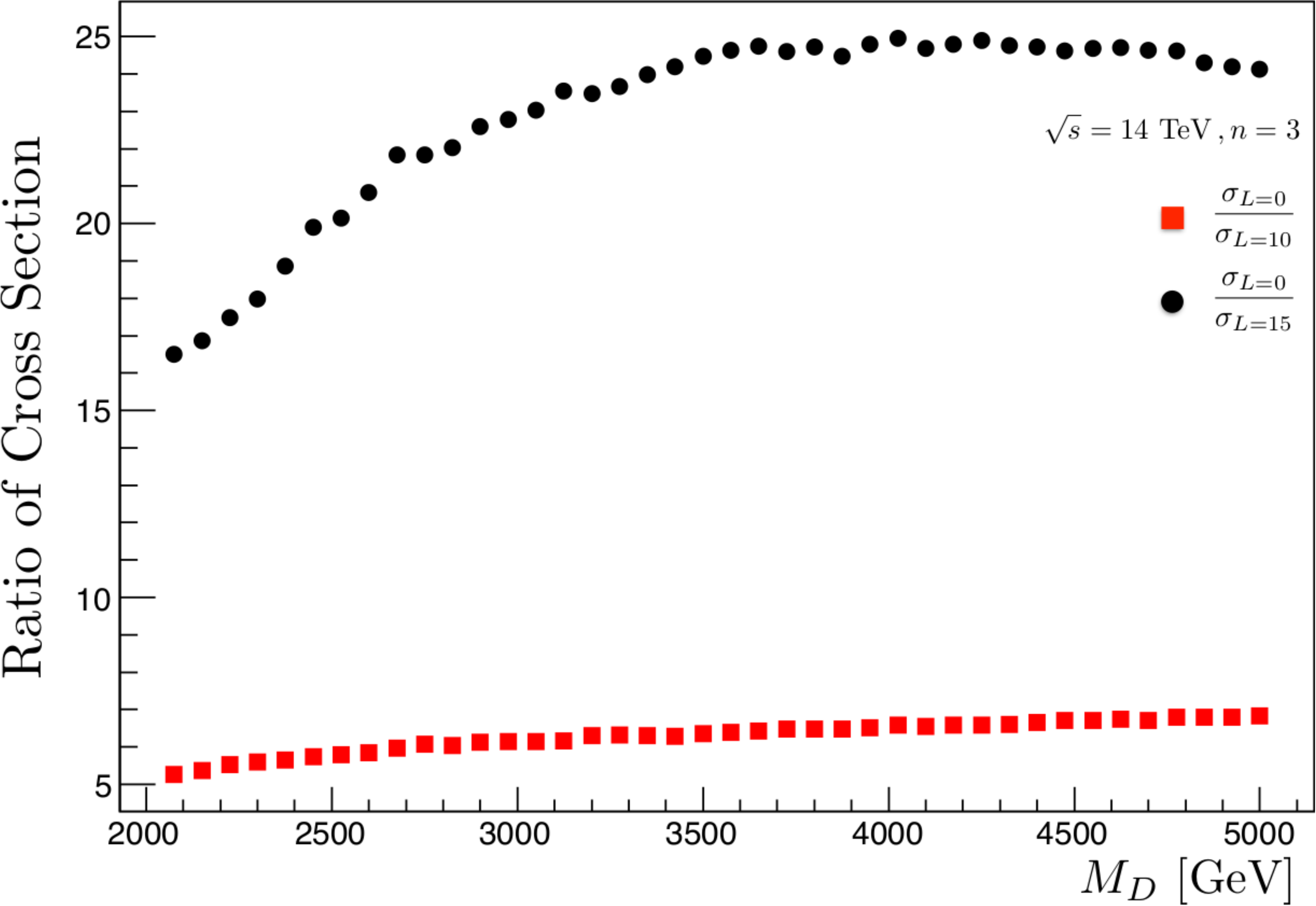}\newline\newline
    \includegraphics[width=10cm]{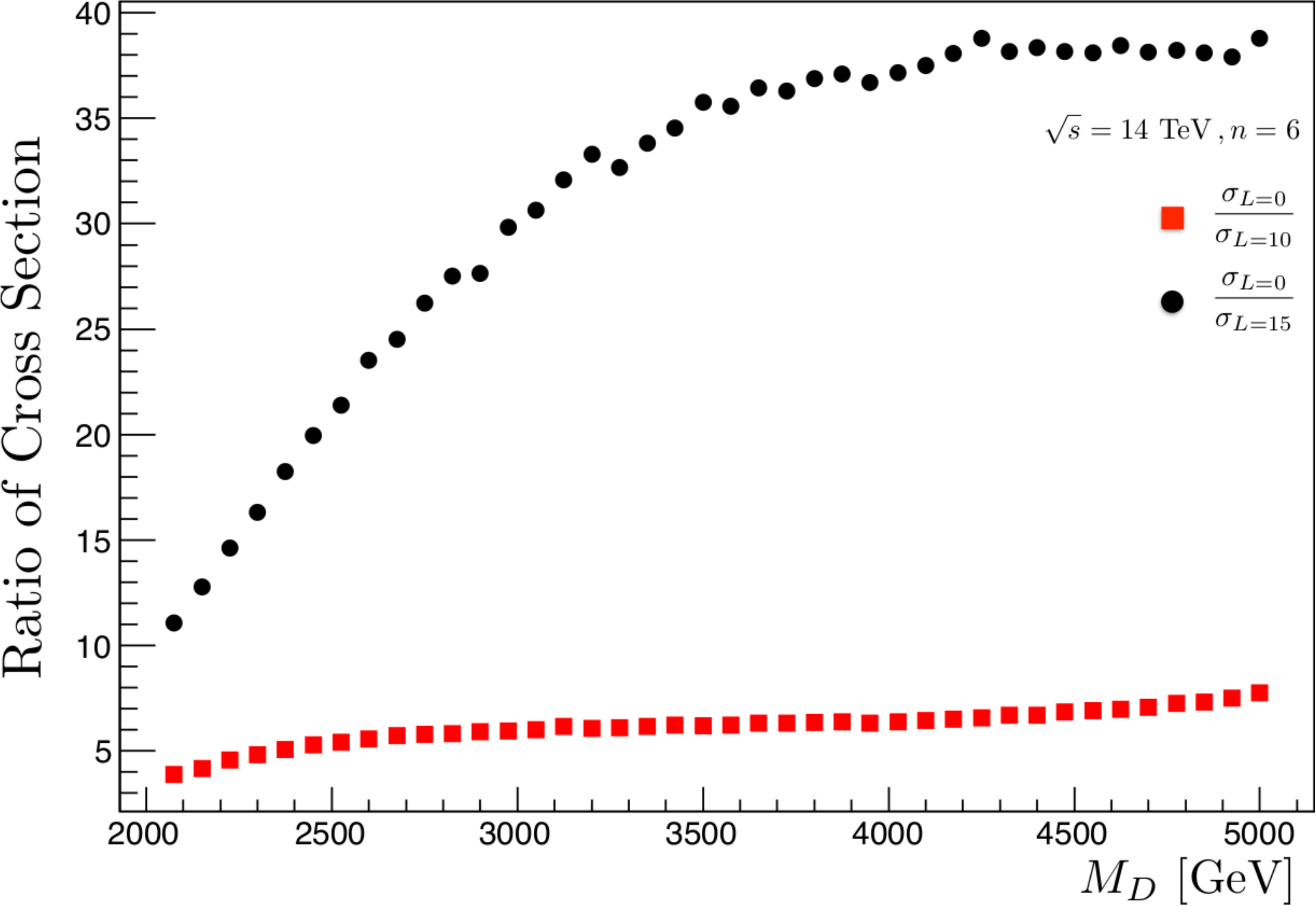}
  \caption{ \label{Fig_13}   Ratio of string ball production for non split case to the string ball production for two-dimensional split fermion model. The first plot shows the ratio for $n=3$ and $g_s=0.1337$. The second plot shows the ratio for $n=6$ and $g_s=0.1589$. \label{Fig_13}}
\end{figure}
\end{center}
\begin{center}
\begin{figure}[htb]
\centering
\hspace{+1.2cm} \includegraphics[width=10cm]{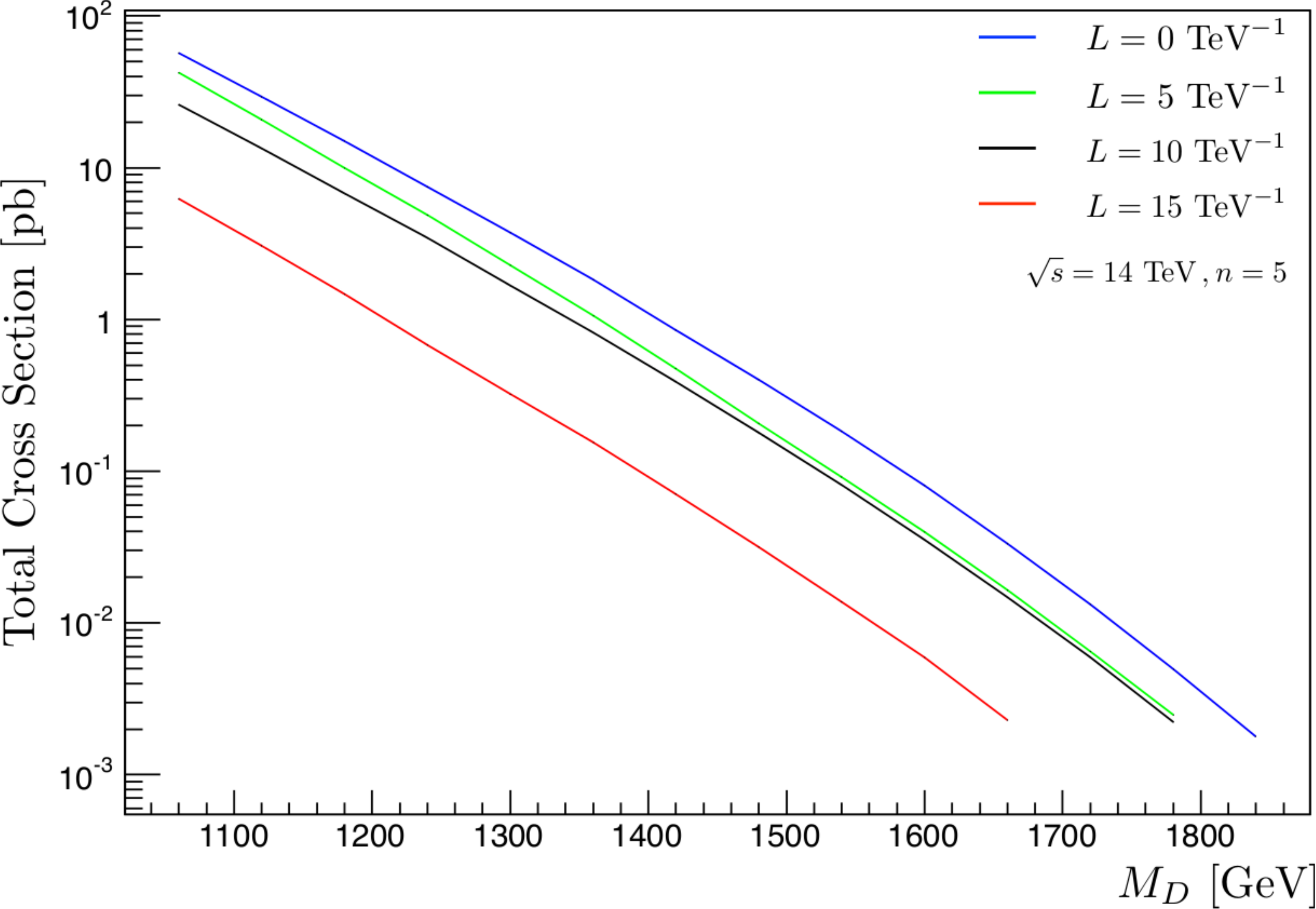}\newline\newline  \includegraphics[width=10cm]{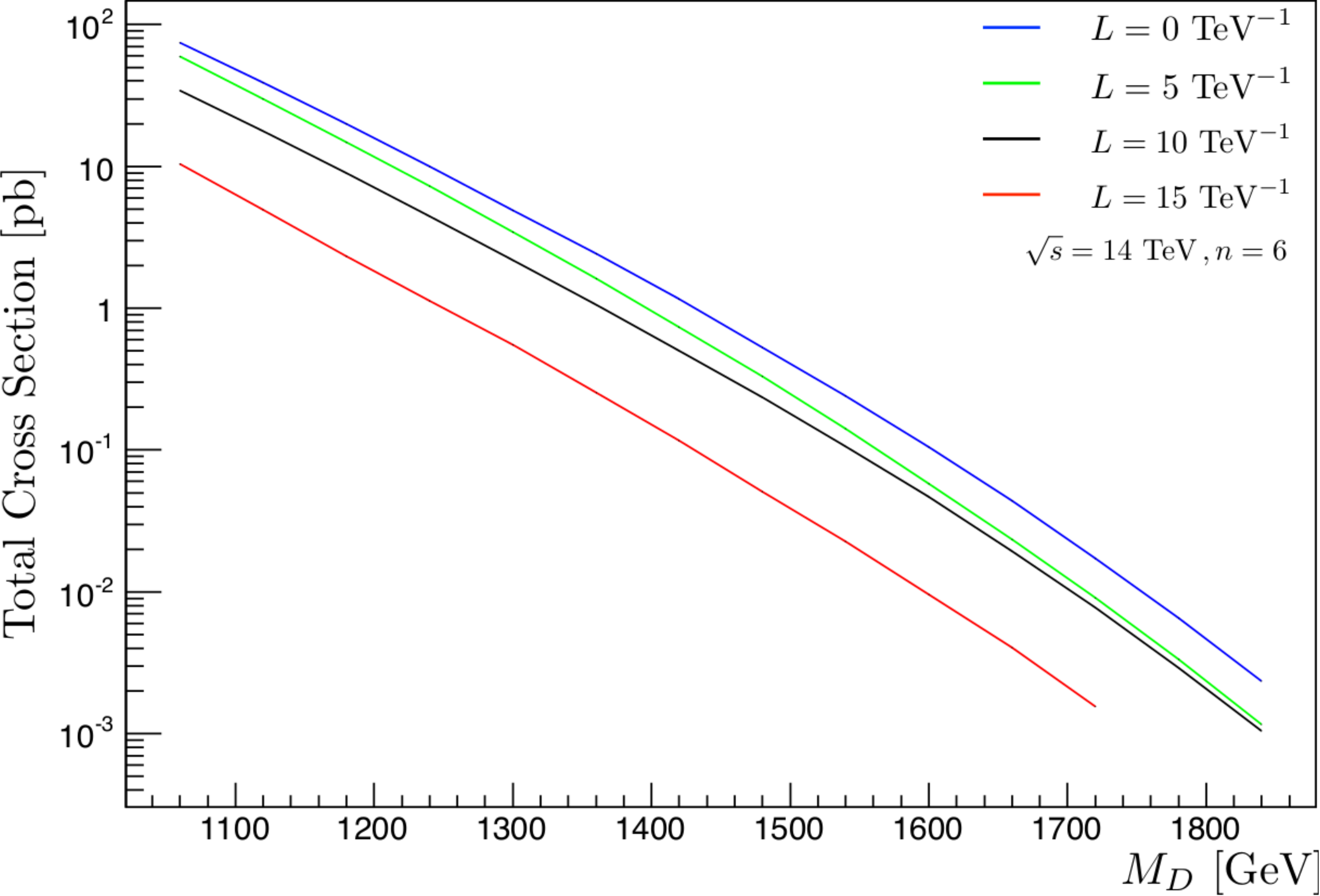}
  \caption{ \label{Fig_14}   The first plot shows the total cross-section of black hole production for $\sqrt{s}=14$ TeV, $n=5$ and L=0, 5, 10, 15 $\text{TeV}^{-1}$. The second plot shows the total cross-section of black hole production for $\sqrt{s}=14$ TeV, $n=6$ and L=0, 5, 10, 15 $\text{TeV}^{-1}$. \label{Fig_14}}
\end{figure}
\end{center}
\begin{figure}[tbp]
\hfill
  \includegraphics[width=10cm]{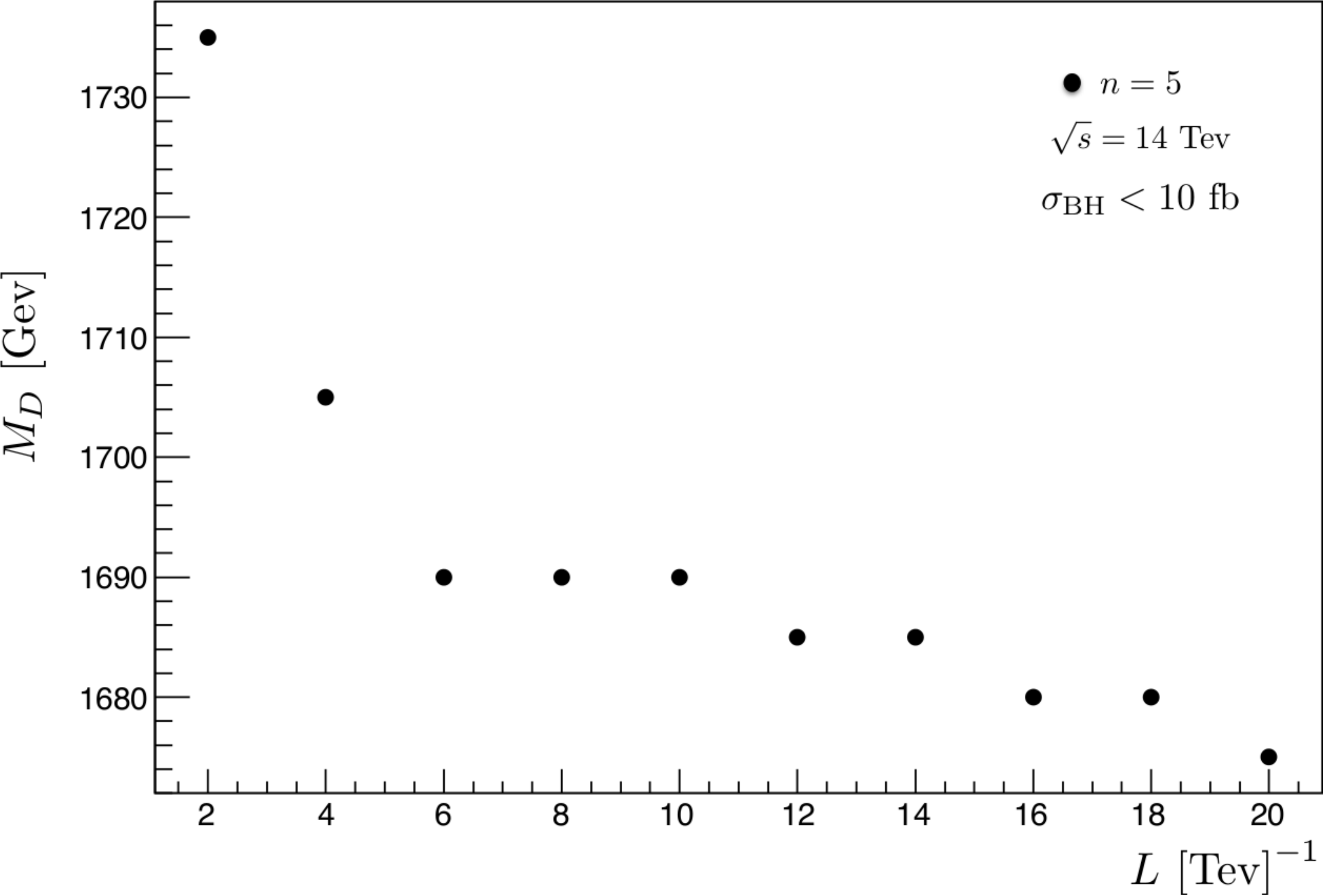}
  \hfill  \, { }
  \caption{ \label{Fig_15}  Threshold for black hole production for $n=5$ and $\sqrt{s}=14$ TeV. In this plot the limit for the cross-section is 10 fb. 
   \label{Fig_15}}
\end{figure}
\begin{figure}[tbp]
\hfill
  \includegraphics[width=10cm]{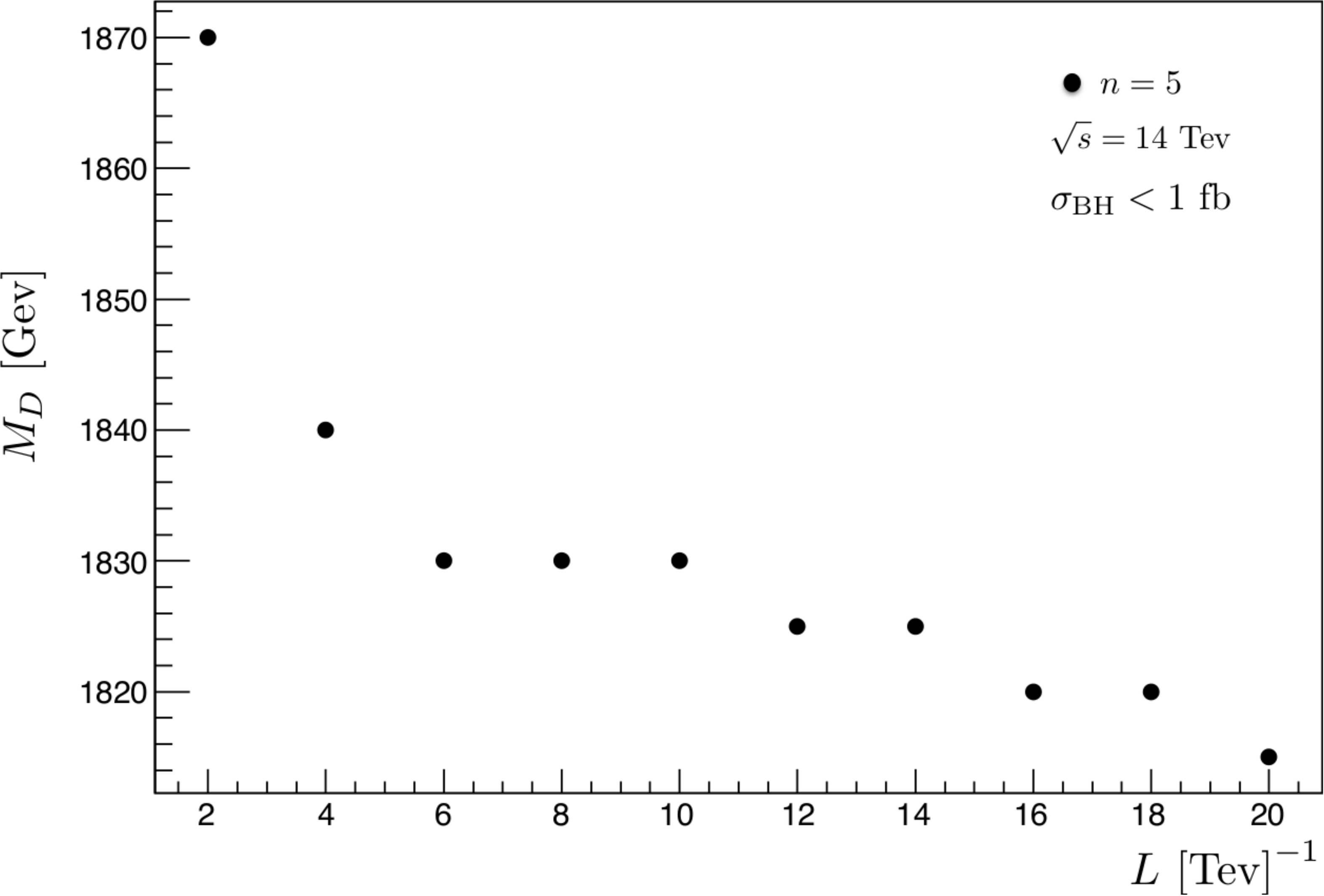}
  \hfill  \, { }
  \caption{ \label{Fig_16}  Threshold for black hole production for $n=5$ and $\sqrt{s}=14$ TeV. In this plot the limit for the cross-section is 1 fb. 
   \label{Fig_16}}
\end{figure}
\begin{center}
\begin{figure}[t]
\centering
\hspace{+1.2cm} \includegraphics[width=10cm]{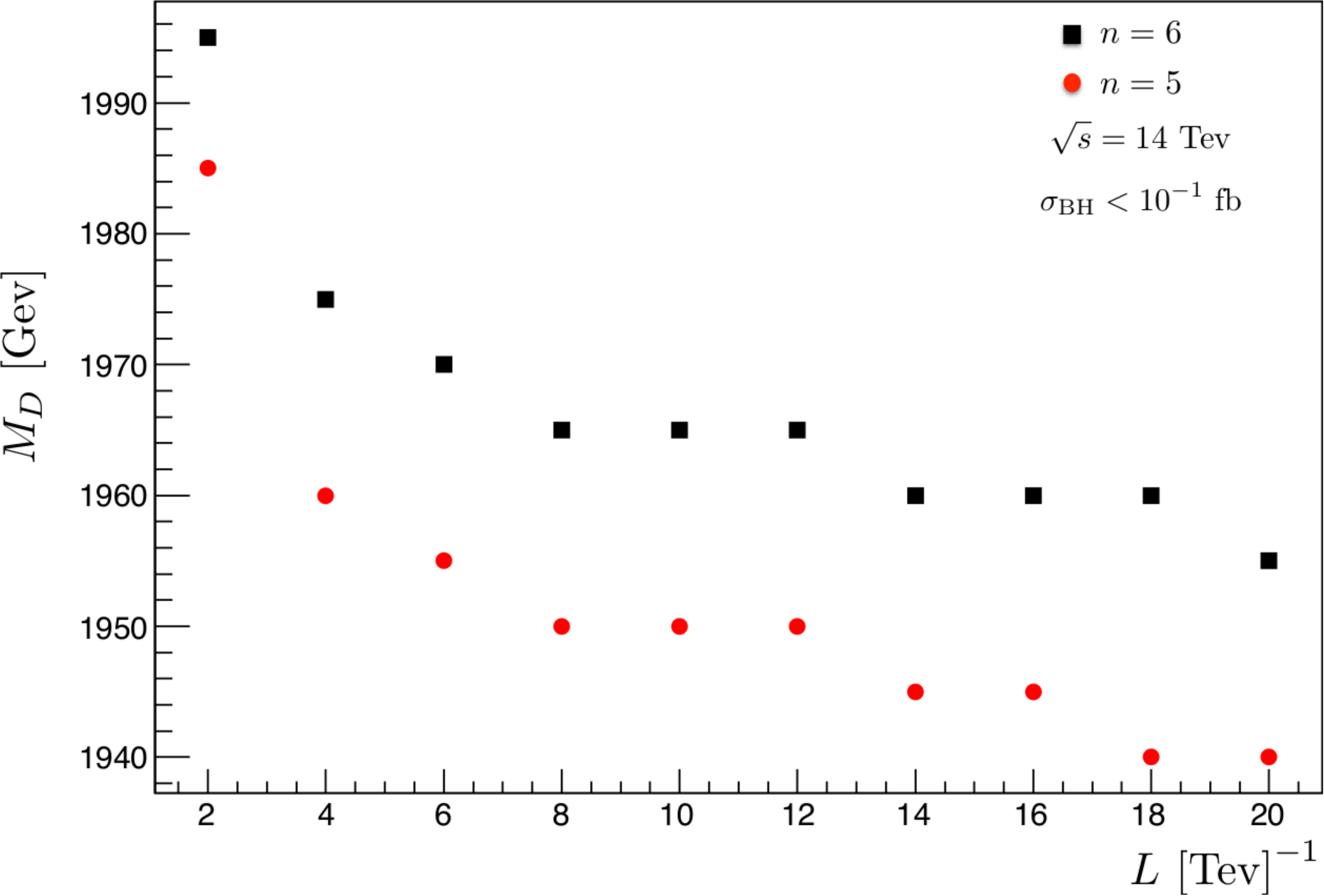}\newline\newline
 \includegraphics[width=10cm]{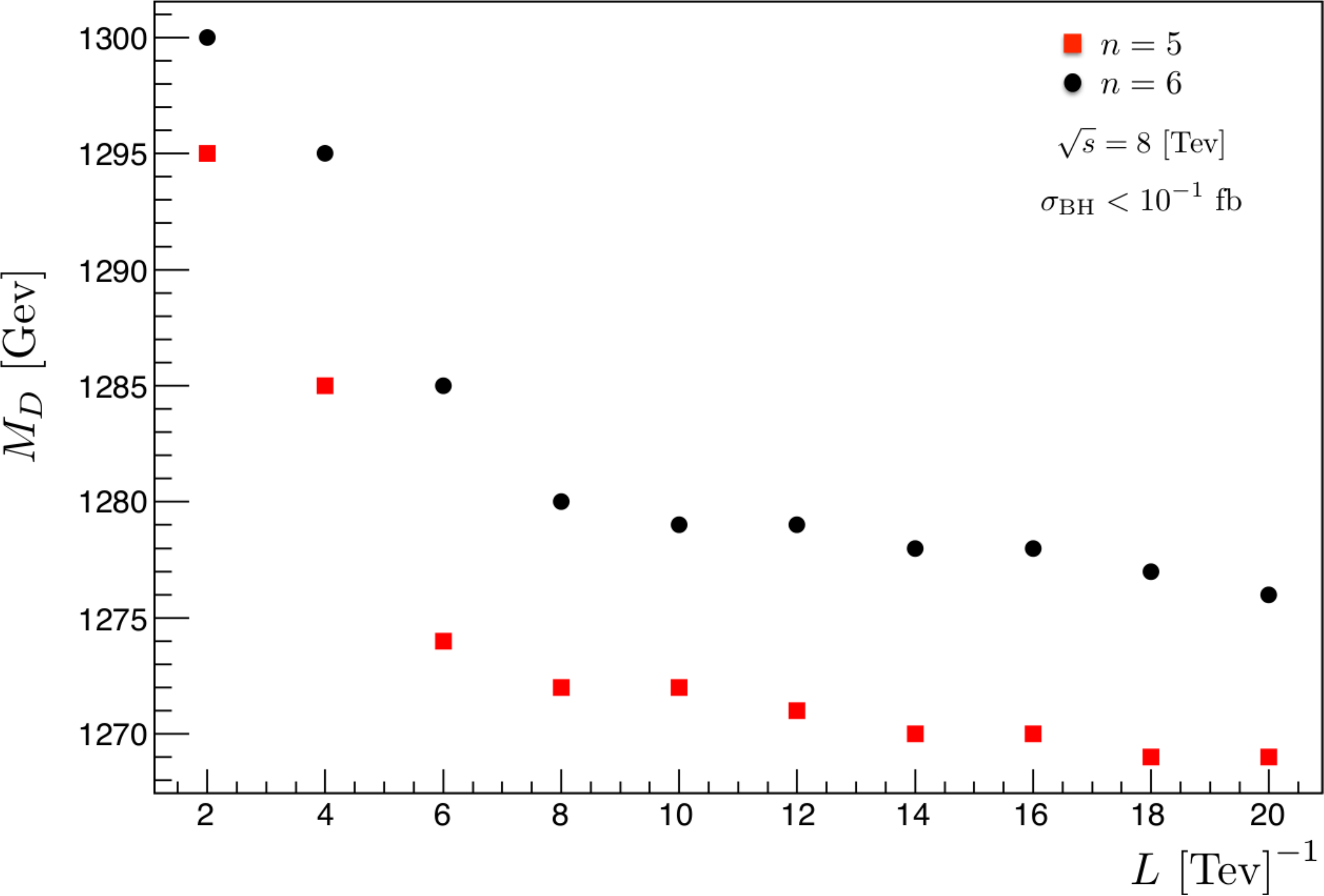}
  \caption{ \label{Fig_17}   The first plot shows the threshold for black hole production for $\sqrt{s}=14$ TeV. In this plot the limit for the cross-section is $10^{-1}$ fb. The second plot shows the threshold for black hole production for $\sqrt{s}=8$ TeV. In this plot the limit for the cross-section is $10^{-1}$ fb. \label{Fig_17}}
\end{figure}
\end{center}
\begin{center}
\begin{figure}[htb]
\centering
\hspace{+1.2cm} \includegraphics[width=10cm]{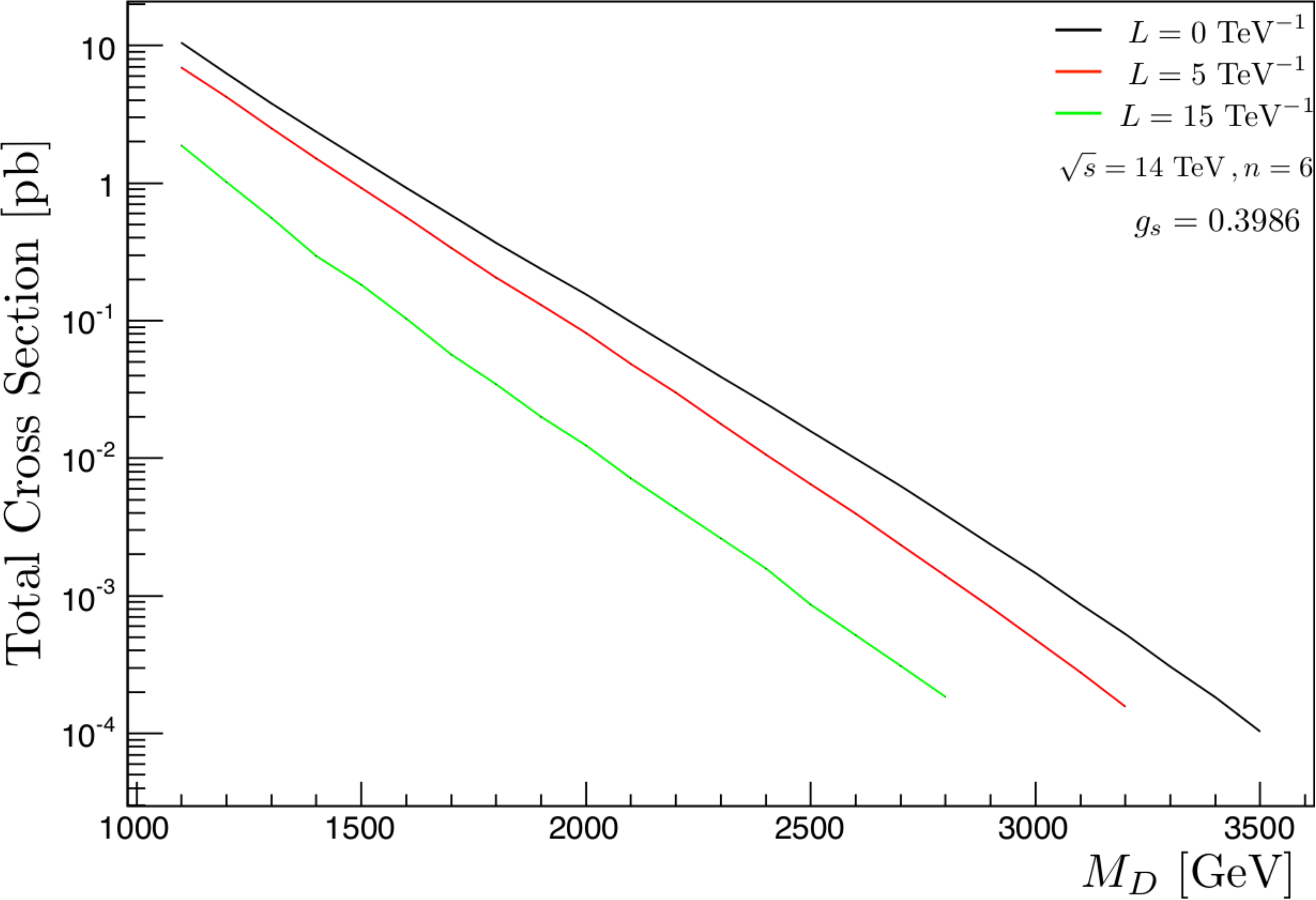}\newline\newline  \includegraphics[width=10cm]{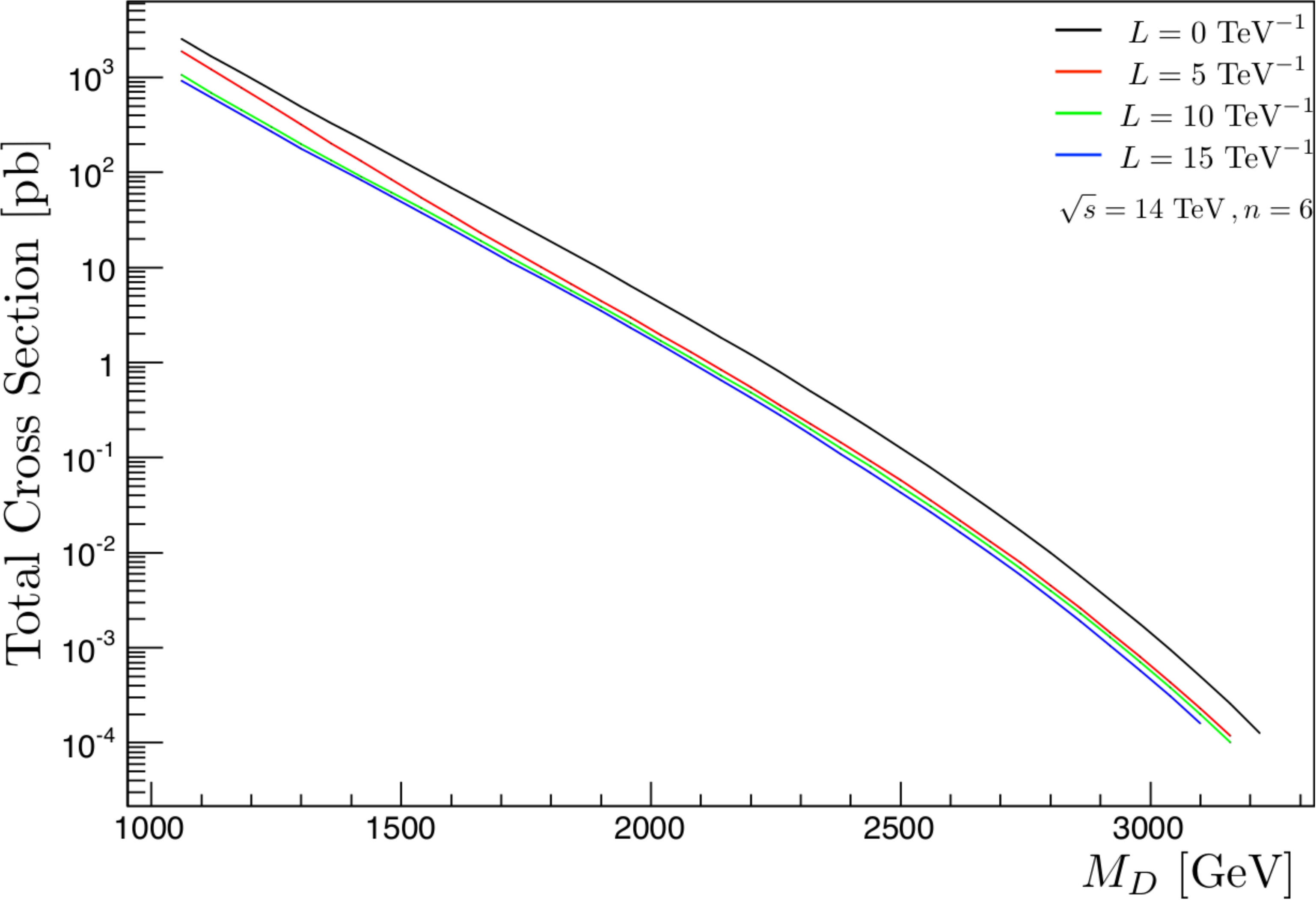}
  \caption{ \label{Fig_18}   The first plot shows the total cross-section of string ball production for $\sqrt{s}=14$ TeV, $n=6$ and L=0, 5, 15 $\text{TeV}^{-1}$. The second plot shows the total cross-section of black hole production for $\sqrt{s}=14$ TeV, $n=6$ and L=0, 5, 10, 15 $\text{TeV}^{-1}$. \label{Fig_18}}
\end{figure}
\end{center}

We will not be able to cover all the range of parameter space, rather we
try to simply illustrate and discuss the sensitivity of the production
cross-section on some of the parameters.
We have used the BlackMax code for our numerical computation, for details see Appendix A. 

\subsection{Discussion} 
We have considered the cross-section for black hole and string ball
production in proton-proton collisions in a two-dimensional
split-fermion TeV-scale gravity model.  

The cross-section of string ball and black hole production decrease as $M_D$
increases.  
In all the following figures, we do not consider values of 
$M_D$, where the cross-section drops below
about $10^{-5}~\text{pb}$. At the integrated luminosity of 20.3 fb$^{-1}$, for the run of the LHC at $\sqrt{s}=8$ TeV \cite{39}, this cross-section corresponds to no events.
In the figures \ref{Fig_1}--\ref{Fig_8}, we consider centre-of-mass energy $\sqrt{s}=8~\text{TeV}$. The figures \ref{Fig_1}--\ref{Fig_7} illustrate the dependence of the cross-section for string ball production versus $M_D$ to many different parameters of the theory and especially to the string coupling constant ($g_s$). In the figures \ref{Fig_1}--\ref{Fig_3}, we can see the cross-section of string ball production for different number of extra dimensions and different values of $L$. In these figures, we take into account two different values of $g_s$, the first one is calculated for parameter $\gamma=0.2$, and the second one is calculated for $\gamma=0.5$. 
The cross-section of string ball production significantly depends on value of $g_s$, therefore, is model dependent drastically. One of the goals of this paper is to investigate the relation between the cross-section and $g_s$. In particular, the range of values of the string coupling constant considered in \cite{35} ($0.02<g_s<0.2$) is investigated. We can see from the graphs that, the cross-section of string ball production for the range of values of the string coupling constant of  $0.02<g_s<0.2$ is really high, even in the case that the centre-of-mass energy is only 8 TeV. By taking into account that we have no signature of string balls in the last run of the LHC \cite{39}, we conclude that $g_s$ cannot be in that range if we want to have a theory of quantum gravity with low fundamental Planck scale. In the figures \ref{Fig_4}--\ref{Fig_6}, the values of string coupling constant that we take into account are the following: $g_s=0.1589$, $g_s=0.2683$, and $g_s=0.3986$. The first two values are calculated for parameter $\gamma=0.2$ and $\gamma=0.5$, respectively. For the last value of $g_s$, we let the value of $\gamma$ to vary with the number of the extra dimensions according to the equation (\ref{eq6}). The value of the width of the brane is constant for each figure and it is equal to $L=0 \, , 5 \, , 15$ $\text{TeV}^{-1}$, respectively. These three figures are for the same number of extra dimensions ($n=6$) and centre-of-mass energy (8 TeV). 

Moreover, from figures \ref{Fig_1}--\ref{Fig_7}, we see that the cross-section drops as the width of split fermion dimension $L$ increases from $L=0$ for non-split fermion model to $L=15~ \text{TeV}^{-1}$ for a ``thick'' two-dimensional split fermion model. In the figure \ref{Fig_7}, we illustrate the cross section of the string ball production for extremely thick split fermion brane.

The last set of plots for the 8 TeV has to do with the cross-section of black hole production. In figure~\ref{Fig_8}, we have considered the cross-section of the black hole production in $n=5$ and $n=6$ extra dimensions at centre-of-mass energy 8~TeV. Furthermore, we vary the number of the extra dimensions and the with of the split fermion brane $L$. Taking into account the current experimental limits on $M_D$ for different number of dimensions, and the fact that the mass of the black hole has to be close to the maximum LHC energy for it to be described by general relativity ($m\geq 5M_D$), a black hole production cross-section is significantly reduced at the LHC.
 
In the figures \ref{Fig_9}--\ref{Fig_14}, we consider centre-of-mass energy to be 14~TeV. The figures~\ref{Fig_9}--\ref{Fig_11} illustrate more the dependence of the string ball cross-section in the split-fermion model and in the non-split-fermion model ($L=0$) to the value of the string coupling constant $g_s$. From the figure \ref{Fig_9}, we can see that as the value of $g_s$ increases the cross-section of string ball production drops faster as $L$ increases. The figures~\ref{Fig_10} and \ref{Fig_11} illustrate the dependence of the cross-section of the string ball production in various numbers of extra dimensions and $L$ for the same value of $g_s$. The number of extra dimension takes the following values $n=3$, 4, 5, 6. In the figure \ref{Fig_12}, we can see that, in the case that the parameter $\gamma$ changes with the number of dimensions according to the equation (\ref{eq6}), the cross-section drops while the number of dimensions increases for the same value of the width of the brane. 

The figure~\ref{Fig_13}, illustrates the difference of the cross-section of the string ball production between the non-split-fermion and the split-fermion model. We see the drop of the cross-section in the case of the two-dimensional split fermion model by calculating the ratio of the cross-section for $L=0$ to $L=10$ and 15 $\text{TeV}^{-1}$. By comparing the graphs in figure \ref{Fig_13}, we can see how the ratio of the cross-section for string ball production in the non-splitting case to the splitting case depends on the number of extra dimensions. However, the ratio increases at least by a factor of two while the width of the brane increases.

In figure \ref{Fig_14}, we illustrate the dependence of the black hole production cross-section in various number of extra dimensions $n=5$, 6 and different value of $L$ ($L=0$, 5, 10, 15 $\text{TeV}^{-1}$). 

The figures~\ref{Fig_15}--\ref{Fig_17}, illustrate the
threshold for the fundamental Planck scale $M_D$, where the production cross-section drops below
$10^{-1}$~fb, 1~fb, and 10~fb for different values of $L$.
Such figures are very helpful for comparing the threshold $M_D$ for
split-fermion to non-split-fermion models as a function of $L$. In these figures, the error in the value of $M_D$ is $\pm 5$ GeV. In figures \ref{Fig_15} and \ref{Fig_16}, we consider $n=5$. In figure \ref{Fig_17}, we consider $n=5$ and $n=6$ and centre-of-mass energy 8~TeV and 14 TeV. From the figure \ref{Fig_17}, we can see that, while the number of extra dimensions increases, the the fundamental Planck scale $M_D$ where the production cross-section drops below $10^{-1}$~fb, decreases. We do not have threshold for the string ball production because it is $g_s$-dependent therefore model dependent. One can choose a specific $g_s$ and create the threshold for the string ball or to do it for the different values of $g_s$. Last but not least in figure \ref{Fig_18}, we consider the cross-section for string ball and black hole production in the case that $M_\mathrm{min}^\mathrm{c}=M_{min}^s=3 M_D$. This last figure illustrates that for $M_\mathrm{min}^\mathrm{c}=3 M_D$ as compared to $M_\mathrm{min}^\mathrm{c}=5 M_D$ black hole production is larger and string ball production will be slightly smaller.
\section{Conclusion}
One of the predictions of the TeV-scale gravity models is the possibility of
black hole production in proton-proton collisions at the LHC. 
In order to make serious predictions, one needs to work within the context of realistic phenomenologically valid models. We consider that, a two-dimensional split fermion model is such a model for a TeV-scale quantum gravity. Since, it is the one that 
produces the correct quark masses and
the mixing angles, as well as the required strength of CP violation
in quark sector. Because of the correspondence between black holes and
string balls, black hole production is dual to the production of a
highly-excited long string state. In this paper, we presented cross-sections for the black hole and string ball production for a TeV-scale gravity model with split
fermions in two dimensions. 
We compared the cross-section for black hole and string ball production
in proton-proton collisions in the non-split-fermion model to a
two-dimensional split-fermion model.  
We conclude with the following comments:
\begin{enumerate}
\item The cross-section of the string ball production is higher than the black hole especially for $g_s$ less than $0.25$. 
\item The cross-section of the black hole or the string ball production drops while the width of the brane increases. The last feature is obvious in the figure \ref{Fig_13}, where we see the drop of the cross-section to be by a factor of at least two with the increase of the width of the brane from $L=10$ to 15 $\text{TeV}^{-1}$. 
\item The cross-section for string ball production in two-dimensional split fermion model reduces more in comparison to black holes. 
\item Black hole is quite hard to be observed at the LHC. In particular, for $\sqrt{s}=14$ TeV and in the case that the fundamental Planck scale is greater than the current experimental limits \cite{51}, black holes cannot be produced at the LHC. In the figure \ref{Fig_14}, we illustrate that in order to have cross-section greater than $10^{-3}$ pb, $M_D$ must be less than or near 1.8 TeV. In the case of centre-of-mass energy 8 TeV, in order to have cross-section greater than $10^{-4}$ pb, $M_D$ must be less than or near 1.3 TeV as it is illustrated in figure \ref{Fig_8}. Note that, the actual value of $M_D$ where the cross-section drops below any value depends on the width of the brane. Such that the thicker the brane, the smaller the value of $M_D$ where the cross-section drops below any given value. 
\item Cross-section for string ball depends significantly on string coupling constant, making it very model dependent. We suggest that using the current data of the LHC at 8 TeV perhaps, one can already rule out some range of the parameter space. According to ATLAS and CMS collaborations results \cite{39,44}, no string balls or black holes have been observed which lead us to conclude, if we still want to believe in string ball model, then, the value of the string coupling constant in six dimensions is close to or larger than $g_s=0.3986$ which we take into account for one of the plots in figures \ref{Fig_4}--\ref{Fig_6}. Even then, the value of cross-section for string ball production is high, unless one considers a proper range of $M_D$, or a very thick brane. 
\item Given the lack of signature of black holes or string balls for the centre-of-mass energy $\sqrt{s}=8$ TeV at the LHC, limits on the value of the fundamental Planck scale can be imposed. These limits depend on the width of split fermion dimension and the value of the string coupling constant. In particular, comparing the black line in figure \ref{Fig_6} and the green line in \ref{Fig_4}, we see that, for the case $L=0~\text{TeV}^{-1}$ in order the cross section to drop below 0.1 fb the value of the fundamental Planck scale of our world should be greater than 2.2 TeV while, for the case $L=15~\text{TeV}^{-1}$, $M_D>1.9$ TeV. 
\item As we can see from figure \ref{Fig_7}, if one wishes to have cross-section for production of string balls less or close to $1$ fb even for $M_D=1$ TeV, one has to consider a two-dimensional split fermion model with extremely thick brane. In such a model, one can explain the lack of signature of string balls at the LHC.
\item From figures \ref{Fig_4}--\ref{Fig_6}, we see that, in six extra dimensions, if $M_D\geq 2.51$ TeV, in the case that, $g_s$ is given by equation (\ref{eq7}), (or when parameter $\gamma$ is given by equation (\ref{eq6})), then, string balls cannot be observed at the LHC for $\sqrt{s}=8$ TeV. This is also valid if $g_s$ is near the value coming from equation (\ref{eq7}).
\end{enumerate}
\acknowledgments

The authors gratefully acknowledge support by the 
DFG Research Training Group 1620 ``Models of Gravity''.
C. T. is grateful to the Natural Sciences and Engineering Research Council of Canada
for its financial support. Moreover, the authors would like to thank Dr. Andrei Zelnikov and Dr. De-Chang Dai for their help and useful advices for the installation and running of  BlackMax code.
\appendix
\section{Appendix}

In this appendix, we give some of the details about the parameter file. There are three categories of variables in the parameter file. 

\begin{enumerate}
\item The variables that, we do not take into account because they do not affect the cross-section for the formation of the black hole but, the evaporation part. 
\item The variables that, we take into account but, they are fixed for all of our figures. 
\begin{itemize}
\item Incoming particles: For this parameter we set the value to be always $1$ since we take into account proton-proton collisions.
\item Definition of $M_{pl}$: Also, here the value is always $1$ because we use the PDG definition of the Planck mass~\cite{54}. 
\item Choose a case: Here we choose the case of tensionless non-rotating black hole. Therefore, the value we consider is $1$ as well.
\item Number of splitting dimensions: The model we consider here is a two-dimensional split fermion model. Thus, the value of this parameter should be equal to $2$.
\item Tension: As we have mentioned we take into account tensionless non-rotating case. Therefore, the value of this parameter is $1.0$.
\item Choose a pdf file: We consider the CTEQ parton distribution functions so, the value should be from $200$ to $240$. We chose this value to be $200$. 
\item Other definition of cross-section: We do not want to have any other definition of the cross-section beside the default one. We set the value of this parameter equal to $0$.
\item Calculate the cross-section according to: We want calculation of the cross-section according to the radius of initial black hole so, we choose this value to be $0$ as well. For this case, the cross-section is calculated in the same way as in ~\cite{53}.
\item Mass, momentum, angular momentum loss factor: We consider only $10\%$ loss of the mass, momentum, and angular momentum. So, the values for these three different parameters are $0.1$. 
\end{itemize}
\item The variables that, we take into account but, they change for every different graphs or for every different run of BlackMax code. 
\begin{itemize}
\item Number of extra dimensions.
\item Maximum mass: In the black hole case we have this upper limit equal to the centre-of-mass energy. In the string ball case we set this limit equal to $5M_D$ unless $5M_D>\sqrt{s}$ where the maximum mass will be the energy of centre-of-mass. Note that, $5M_D=M_s / {g_s}^{2}$. If one wants to investigate the cross-section of black hole and string ball production together then, this limit must be set equal to the centre-of-mass energy ($\sqrt{s}$). We do not consider this case in this paper.
\item Minimum mass: In the black hole case this limit is equal to $5M_D$. In the string ball case we consider the minimum mass to be $3M_s$. If one wants to investigate the cross-section of black hole and string ball production together then, this limit must be set equal to $3M_s$.
\item centre-of-mass energy of incoming particle.
\item Include string ball.
\item $M_{pl}$: This parameter must be entered in units of GeV.
\item String scale $(M_s)$.
\item String coupling $(g_s)$.
\item Split fermion width and location: The split-fermion width
$w$ is defined in accordance to the Gaussian wave function of the
form 

\begin{equation}\label{eq11}
\exp\left( -\frac{2x^2}{w^2} \right)\, .
\end{equation}

Thus, if one uses the form $\exp\left( -x^2\mu^2\right)$ for the Gaussian
profile in accordance to Eq.~(\ref{eq1}) then, the split-fermion width in
BlackMax is $w = \sqrt{2}/\mu$, which must be entered in unit of
$1/M_D$,  
\begin{equation}\label{eq13}
w = \frac{\sqrt{2} L}{c}\, .
\end{equation}
The location of the quarks is given in the table \ref{tab1} in units of the $\mu^{-1}$, but we need to enter them in units of $L/30$ in the BlackMax code. If we take into account that,
\be
\mu^{-1}=\frac{L}{c} \, , 
\ee 
and consider $c=30$ we end up to enter the location of the particles as they are in the table \ref{tab1}. In the case that, $c\neq 30$ one takes a dimensionless number from table~\ref{tab1} and multiplies it by $30/c$ in order to enter the correct locations of quarks into the parameter file.  
\item Extra dimension size: A very tricky part of the parameter file is the value of the ``Extra dimension size''. We should stress here that it is different from the ``size of brane'' parameter. Extra dimension size is a parameter that concerns someone who works on the no splitting case. For example, if the number of the splitting dimensions is 2 and total number of extra dimensions is 4. Then, the first two space dimensions use the width of the split brane. The other two directions use this parameter as its brane size. If there is no splitting brane, then this parameter is the only functioning parameter.  We keep the value of this parameter fixed and equal to $1$ in unit of $1/M_D$. The parameter $L$ must be entered in the line corresponding to the extra-dimension size in units of $1/M_D$. In our case, we first choose $M_D$ and $L$. Then if, for example, $M_D = a$~GeV, and $L^{-1} = b$~TeV, one has to enter in the parameter file

\begin{equation}\label{eq12}
L = \frac{a}{1000b} \frac{1}{M_D}\, .
\end{equation}.
\end{itemize}
\end{enumerate}



\end{document}